\documentclass[a4paper,fleqn,usenatbib,useAMS]{mpazh}
\pdfoutput=1

\usepackage{hyperref}

\usepackage[utf8]{inputenc}
\usepackage{graphicx}
\usepackage{grffile}
\usepackage{natbib}

\usepackage{floatrow}
\usepackage{caption}
\usepackage{longtable}
\usepackage{amssymb}

\usepackage{multirow}
\usepackage{booktabs}
\usepackage{amsmath}

\usepackage{lmodern}

\usepackage[toc]{appendix}

\usepackage{multirow}

\usepackage[T1]{fontenc}

\newcommand{\apspr}{Astrophysics\ and\ Space\ Physics\ Research}

\def\ps1{\emph{Pan-STARRS1}}

\def\srg{\textit{SRG}}
\def\art{ART-XC}
\def\ero{eROSITA}
\def\srge{\textit{SRG/eROSITA}}
\def\rosat{\textit{ROSAT}}

\def\xmm{\textit{XMM-Newton}}
\def\xmmshort{\textit{XMM}}

\def\einstein{\textit{Einstein}}

\def\nh{N_{\rm H}}
\def\lx{L_{\rm X}}

\def\fwhm{FWHM}
\def\fwhmmes{FWHM_{\rm mes}}
\def\fwhmres{FWHM_{\rm res}}

\def\ra{R_{\rm a}}
\def\re{R_{\rm e}}

\def\azt{AZT-33IK}
\def\rtt{RTT-150}

\def\ha{\rm $H\alpha$}
\def\hb{\rm $H\beta$}
\def\hd{\rm $H\delta$}
\def\hg{\rm $H\gamma$}
\def\o3hb{\rm $\lg({}[OIII]\lambda5007/H\beta)$}
\def\n2ha{\rm $\lg({}[NII]\lambda6584/H\alpha)$}

\def\ass{ARTSS1-5}

\title{New Active Galactic Nuclei Detected by the \art\ and \ero\ Telescopes during the First Five \srg\ All-Sky X-ray Surveys. Part 2}
\author[Uskov et. al.]{
	G.S. Uskov,
			\thanks{\href{mailto:uskov@cosmos.ru}{\nolinkurl{uskov@cosmos.ru}}}
				$^{1}$
		S. Yu. Sazonov,
				$^{1}$
    	I. A. Zaznobin,
				$^{1}$
        M. R. Gilfanov,
				$^{1,2}$
        \newauthor
    	R. A. Burenin,
				$^{1}$
        E. V. Filippova,
				$^{1}$
        P. S. Medvedev,
    				$^{1}$
        A. V. Moskaleva,
                    $^{1}$
        \newauthor
        R. A. Sunyaev,
				$^{1,2}$
    	R. A. Krivonos,
				$^{1}$
		M. V. Eselevich,
				$^{3}$
	\\
			$^{1}$Space Research Institute, Russian Academy of Sciences, Moscow,
117997 Russia\\
			$^{2}$Max Planck Institut for Astrophysik, Karl-Schwarzschild-Str. 1,
85741 Garching, Germany\\
			$^{3}$Institute of Solar--Terrestrial Physics, Russian Academy of
Sciences, Siberian Branch, Irkutsk, 664033 Russia}

\date{Accepted XXX. Received YYY; in original form ZZZ}

\pubyear{2024}

\begin{document}
\label{firstpage}
\pagerange{\pageref{firstpage}--\pageref{lastpage}}
\maketitle

\begin{abstract}

We present the results of our identification of 11 X-ray sources detected on the half of the sky $0^\circ<l<180^\circ$ in the 4--12~keV energy band on the combined map of the first five all-sky surveys with the Mikhail Pavlinsky \art\ telescope onboard the \srg\ observatory. All these sources were also detected by the \srge\ telescope in the 0.2--8~keV energy band, whose data have allowed us to improve their positions and to investigate their X-ray spectra. Five of them have been detected in X-rays for the first time, while the remaining ones have already been known previously, but their
nature has remained unknown. We have taken optical spectra for nine sources with the 1.6-m AZT-33IK telescope at the Sayan Observatory (the Institute of Solar–Terrestrial Physics, the Siberian Branch of the Russian Academy of Sciences); for two more objects we have analyzed the archival spectra from SDSS and the 6dF survey. The objects are classified as Seyfert galaxies (seven Sy1, three Sy1.9, and one Sy2) at redshifts $z=0.029$--0.258. Our analysis of the X-ray spectra has revealed a noticeable intrinsic absorption ($\nh\sim 10^{22}$\,cm$^{-2}$) in two of the four Seyfert 2 galaxies (Sy1.9--2). The spectrum of one more of them (SRGA\,J000132.9+240237) cannot be described within the model of an absorbed Comptonization continuum, which may point to a strong absorption and a significant contribution of the reflected radiation. However, the available \srg\ all-sky survey data are not enough to obtain reliable constraints on the absorption column density in this object, which is also interesting in that it is radio loud. Longer X-ray observations are required to refine the physical properties of this active galactic nucleus. 

{\it Keywords}:  active galactic nuclei, sky surveys, optical observations, redshifts, X-ray observations

\end{abstract}

\section{INTRODUCTION}

From December 2019 to March 2022 the \ero\ \citep{pred21} and Mikhail Pavlinsky \art\ \citep{Pavlinsky_2021_art} X-ray telescopes onboard the \srg\ observatory \citep{srg} conducted four full all-sky surveys and about 40\% of the fifth all-sky survey. Subsequently, the \art\ telescope conducted a deep survey of the plane and central part of the Galaxy for one and a half years and recently (in October 2023) resumed the all-sky scanning \footnote{The \ero\ telescope has been dormant since February 26, 2022.}.

As data are accumulated and processed, the catalogs of sources detected by the \art\ and \ero\ telescopes during the all-sky survey are regularly updated. The first official catalog of sources detected by the \art\ telescope was produced from the data of the first two surveys \citep{artsurvey}, while recently the second official catalog of \art\ sources that includes the data of the first four and incomplete fifth surveys has been published (\ass, \citealt{sazonov2024}). The 4--12~keV energy band is used to detect the sources in the \srg/\art\ all-sky survey. The operating energy band of the \ero\ telescope is 0.2--8~keV, while the sources are detected and the \ero\ catalogs are constructed in several subbands of this energy band.

Because of the shape of the typical spectra for X-ray sources, which, as a rule, fall off with energy above $\sim 2$--3~keV as a power or even exponential law, and in view of the smaller, for objective reasons, effective area of the X-ray telescopes operating at energies $\sim 10$~keV, the numbers of sources in the \ero\ and \art\ catalogs differ significantly: the \ero\ catalog in the 0.2--2.3~keV energy band in the entire sky includes $\sim 4$ million sources, while the \art\ telescope in the 4--12~keV energy band includes about 1500 sources. The relative compactness of the \art\ catalog allows one to set the task of identifying and individually studying all of the sources included in it (or those that have been unknown previously). For the \ero\ catalog this task would require unrealistic efforts and, on the other hand, is not necessary, since catalogs of such a large size are investigated by other methods.

This paper enters into the series of publications devoted to identifying the sources detected during the \art\ all-sky survey. All of the sources being studied in this paper, as well as in other papers of this series, are also present in the catalog of the \ero\ all-sky survey. They drew our attention owing to their presence in the \art\ catalog, which could be indicative of their high brightness or the hardness of their spectrum. The \ero\ data have allowed the position error circles of these sources to be refined and have provided spectral information in the 0.2--8~keV energy band. Most of the sources being analyzed are sufficiently bright for the \ero\ telescope, which has allowed the behavior of their spectra in the soft X-ray band to be studied in detail.

The work on the optical identification of the sources being detected by the \art\ telescope was begun immediately after the beginning of the all-sky survey. This spectroscopic program is being conducted with the 1.6-m \azt\ telescope at the Sayan Observatory (the Institute of Solar-Terrestrial Physics, the Siberian Branch of the Russian Academy of Sciences) and the 1.5-m Russian–Turkish telescope (\rtt) at the T\"{U}BITAK National Observatory incorporated into the \srg\ ground support complex. The targets of our observations are both the X-ray sources being discovered by the \art\ telescope and the previously known X-ray sources whose nature has still remained unknown. As this program is implemented, we have already managed to identify and classify more than 50 active galactic nuclei (AGNs) and several cataclysmic variables (CVs). Some of our results were presented in \citet{zaznobin21, zaznobin22cv} and \citet{uskov22, uskov23}. In addition, during the \srg/\art\ sky survey several X-ray binaries with neutron stars and black holes were discovered and subsequently identified \citep{lutovinov22,mereminsky22}.

In this paper we present the results of our optical identification and classification of another 11 AGNs from the \ass\ catalog. All of them are on the half of the sky $0^\circ<l<180^\circ$ and were detected during the \srg\ all-sky survey not only by \art\ but also by \ero\. This allowed us to construct the X-ray spectra of these sources in a wide energy range, 0.2--20~keV, and to measure their intrinsic absorption. To calculate the luminosities, we use the model of a flat Universe with parameters $H_0=70$~km~s$^{-1}$Mpc$^{-1}$, $\Omega_m = 0.3$.

\section{THE SAMPLE OF OBJECTS}
\label{s:sample}

\begin{table*}
\caption{The sample of X-ray sources}
\label{tab:sources}
\centering
\renewcommand{\arraystretch}{1.5}
\begin{tabular}{rcccccl}
\toprule
№ & \art\ source &  $\ra$ & $F_{\rm A}^{\rm 4-12}$ & \ero\ source & $\re$   & Discovered \\
\midrule
1  & SRGA\,J000132.9$+$240237 & 21.1 & $4.0\pm1.4$ & SRGe\,J000132.4$+$240229 & 8.5  & \srg\\
2  & SRGA\,J001059.5$+$424341 & 19.6 & $2.4\pm0.8$ & SRGe\,J001059.5$+$424351 & 2.4 &  \srg\\
3  & SRGA\,J023800.1$+$193818 & 26.7 & $3.3\pm1.2$ & SRGe\,J023800.0$+$193811 & 2.3 &  \rosat\\
4  & SRGA\,J025900.3$+$502958 & 24.1 & $3.4\pm1.1$ & SRGe\,J025901.0$+$503013 & 2.4 &  \srg\\
5  & SRGA\,J040335.6$+$472440 & 23.3 & $3.2\pm1.1$ & SRGe\,J040336.4$+$472439 & 5.0 &  \rosat\\
6  & SRGA\,J165143.2$+$532539 & 15.8 & $1.4\pm0.4$ & SRGe\,J165144.1$+$532539 & 2.2  & \einstein\\
7  & SRGA\,J181749.5$+$234311 & 17.1 & $5.8\pm1.6$ & SRGe\,J181749.1$+$234313 & 3.0  & \srg\\
8  & SRGA\,J191628.1$+$711619 & 17.1 & $1.1\pm0.3$ & SRGe\,J191629.4$+$711614 & 2.1  & \rosat\\
9  & SRGA\,J194412.5$-$243619 & 20.0 & $5.7\pm1.9$ & SRGe\,J194412.5$-$243623 & 3.4 & \xmmshort\\
10 & SRGA\,J195226.6$+$380011 & 19.0 & $4.3\pm1.4$ & SRGe\,J195225.4$+$380028 & 2.5  & \rosat\\
11 & SRGA\,J201633.2$+$705525 & 18.8 & $1.1\pm0.3$ & SRGe\,J201632.4$+$705525 & 2.3  & \srg\\
\bottomrule
\end{tabular}
\begin{flushleft}
    Column 1: the ordinal source number in the sample being studied; 
    column 2: the source name in the \ass\ catalog; 
    column 3: the radius of the 98\% \art\ position error circle in arcsec; column 4: the 4--12~keV flux from the \art\ data in units of $10^{-12}$~erg~s$^{-1}$~cm$^{-2}$; 
    column 5: the source name in the \ero\ catalog; 
    column 6: the radius of the 98\% \ero\ position error circle in arcsec; 
    and column 7: the orbital observatory that detected the X-ray source for the first time.
\end{flushleft}
\end{table*}

\begin{table*}
\caption{Multiwavelength properties of the objects}
\label{tab:multiwave}
\centering
\renewcommand{\arraystretch}{1.5}
\begin{tabular}{rlrrll}
\toprule
№ & \multicolumn{1}{c}{Counterpart} & \multicolumn{1}{c}{$\alpha$} & \multicolumn{1}{c}{$\delta$} & \multicolumn{1}{c}{$W1-W2$} & \multicolumn{1}{c}{Radio} \\
\midrule
1  & 2MASX\,J00013232+2402304 = NVSS\,J000132+240231 & 0.38474 & 24.04177 & $0.86\pm0.03$ & $359\pm 11$ (1.4\,GHz) \\
2  & WISEA\,J001059.72+424352.8 & 2.74883 & 42.73133 & $0.71\pm 0.05$ & \\
3  & 2MASS\,J02375999+1938118 & 39.50000 & 19.63661 & $0.11\pm 0.03$ & \\
4  & LEDA\,2374943 = VLASS1QLCIR\,J025900.99+503014.6 & 44.75421 & 50.50408 & $0.79\pm0.03$ & $1.6\pm0.3$ (2--4\,GHz)\\
5  & 2MASS\,J04033641+4724383 & 60.90171 & 47.41070 & $0.94\pm 0.03$  & \\
6  & SBS\,1650+535 = NVSS\,165143+532538 & 252.93225 & 53.42772 & $0.75\pm0.03$ & $3.9\pm 0.4$ (1.4\,GHz)\\
7  & LEDA\,1692433 = NVSS\,181749+234313 & 274.45412 & 23.72019 & $0.65\pm 0.03$ & $8.4\pm 0.5$ (1.4\,GHz)\\
8  & WISEA\,J191629.25+711616.4 & 289.12188 & 71.27122 & $0.59\pm0.03$ & \\
9  & 2MASX\,J19441243$-$2436217 = NVSS\,J194412$-$243622& 296.05168 & $-$24.60590 & $1.06\pm0.03$ & $ 5.9\pm 0.5$ (1.4\,GHz)\\
10 & 2MASS\,J19522509+3800269 & 298.10454 & 38.00748 & $0.88\pm 0.03$ & \\
11 & WISEA\,J201632.61+705527.2 & 304.13587 & 70.92422 & $1.04\pm 0.03$ & \\
\bottomrule
\end{tabular}
\begin{flushleft}
    Column 1: the ordinal source number in the sample being studied; 
    column 2: the optical/IR/radio counterpart; 
    columns 3 and 4: the coordinates of the optical counterpart (J2000.0); column 5: the IR color; 
    and column 6: the spectral radio flux density in Jy.
\end{flushleft}
\end{table*}

Objects from the catalog of sources detected in the 4--12~keV energy band on the combined map of the first $\sim 4.4$ \srg\/\art\ sky surveys (from December 12, 2019, to March 7, 2022) (the \ass\ catalog, \citealt{sazonov2024}) constituted the sample. We considered only the sources located on the half of the sky $0^\circ<l<180^\circ$ for which we also have \srg/\ero\ data. The sample includes a total of 11 AGN candidates.

For all objects Table~\ref{tab:sources} gives the X-ray source names in the \art\ and \ero\ catalogs, the corresponding radii of the position error circles at 98\% confidence ($\ra$ and $\re$), the 4--12~keV flux from the \art\ data, and the name of the orbital observatory that detected the source in X-rays for the first time. Five sources have been detected for the first time in X-rays with the \art\ and \ero\ telescopes onboard the \srg\ observatory.

\begin{figure*}
  \centering
  \includegraphics[width=0.25\columnwidth]{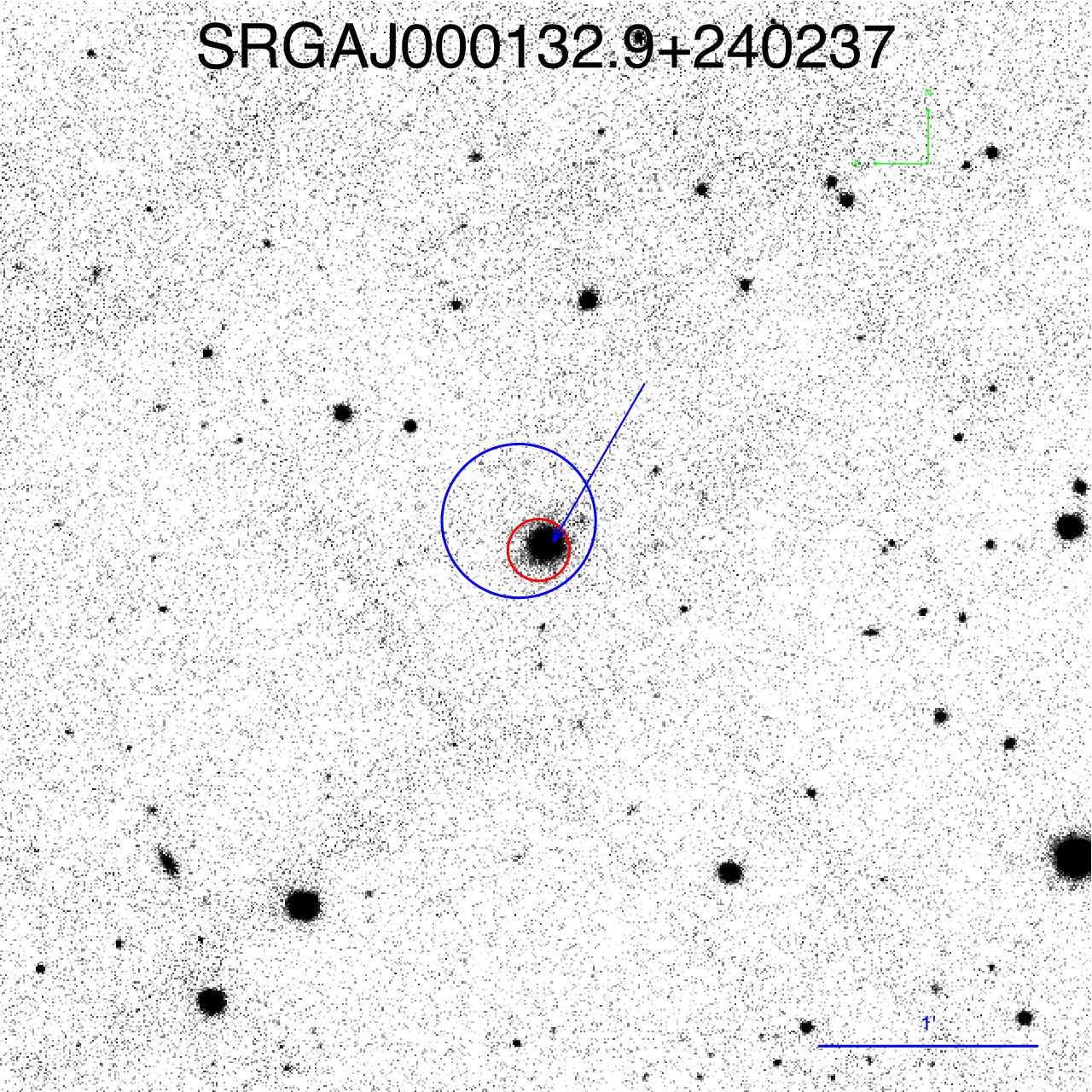}
  \includegraphics[width=0.25\columnwidth]{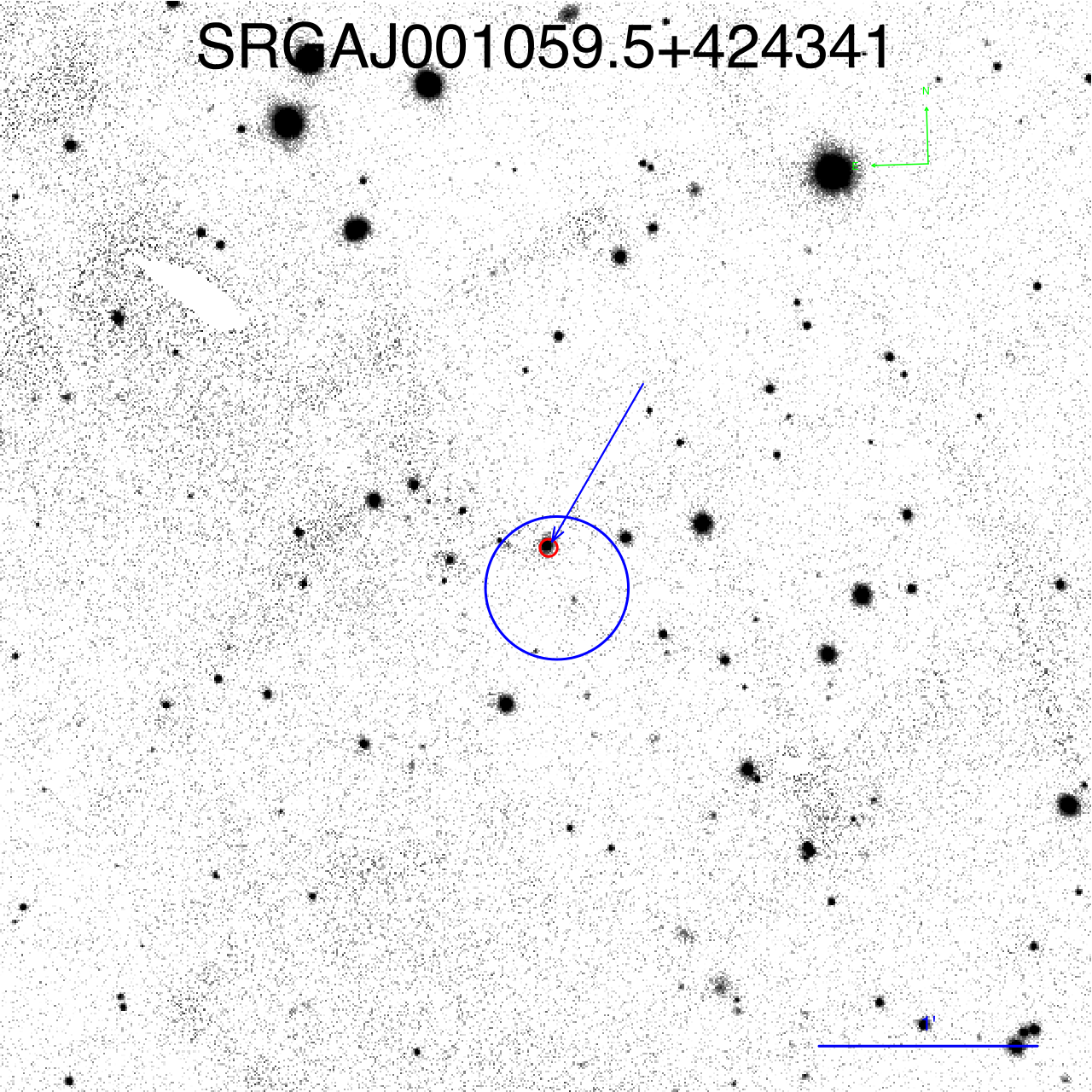}
  \includegraphics[width=0.25\columnwidth]{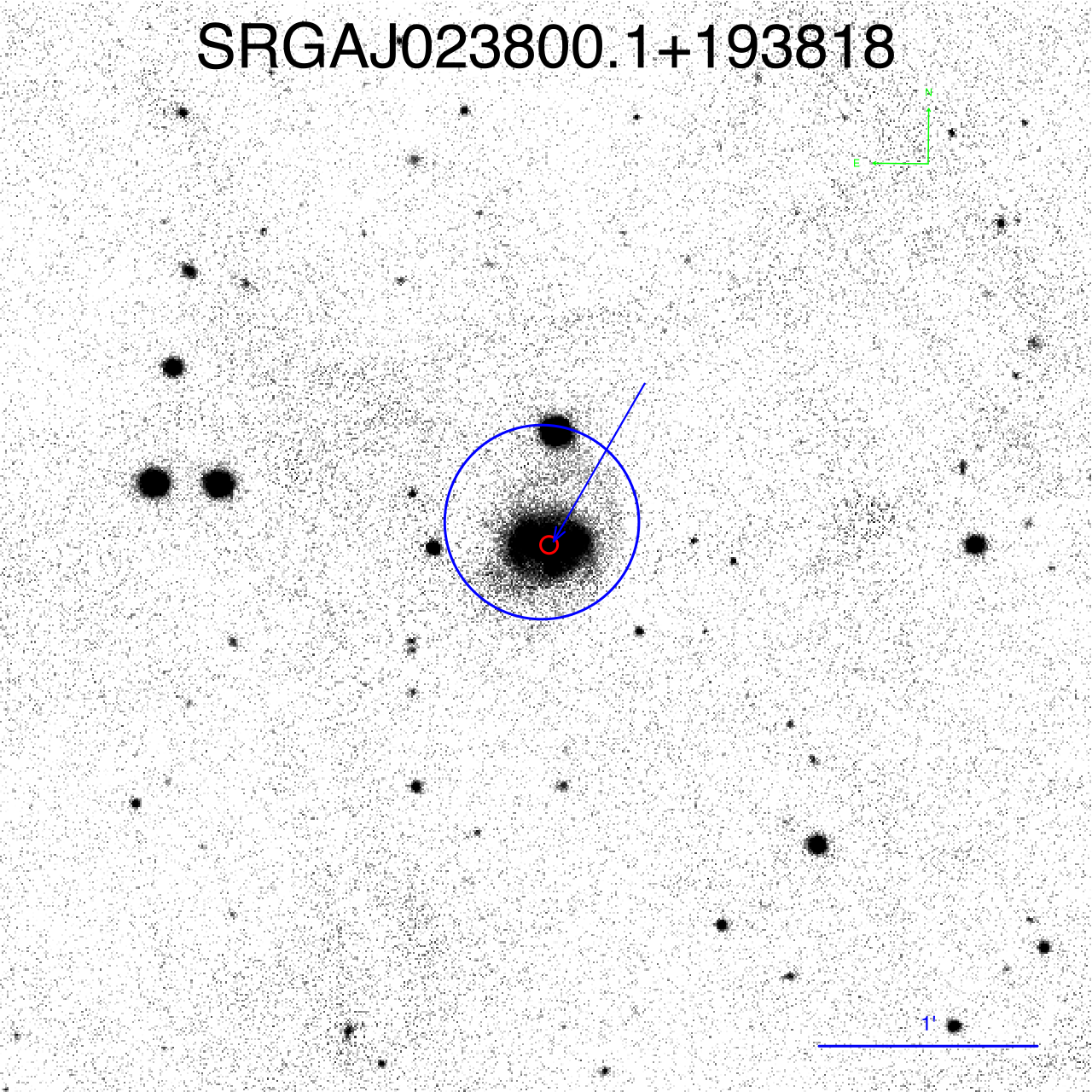}
  \includegraphics[width=0.25\columnwidth]{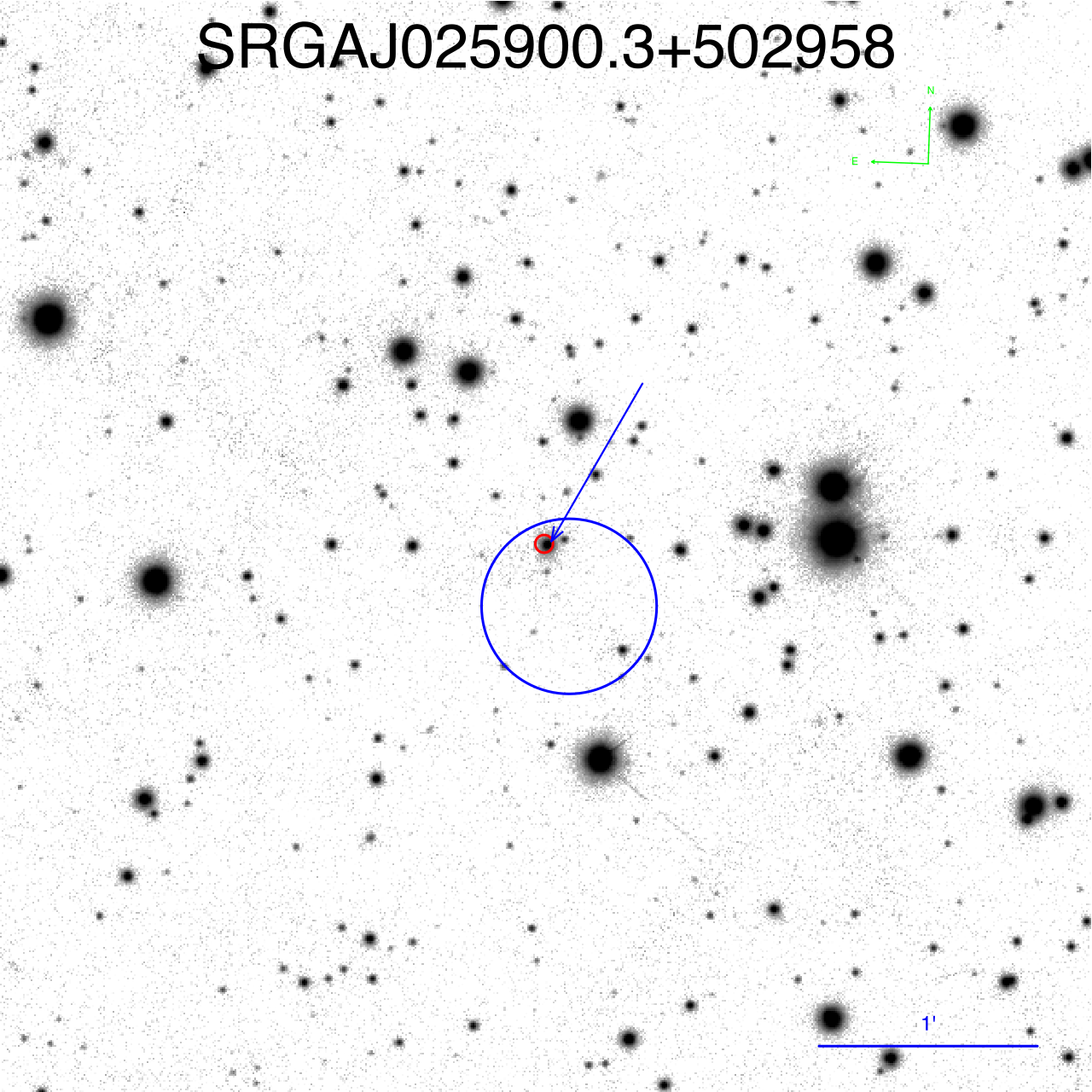}
  \includegraphics[width=0.25\columnwidth]{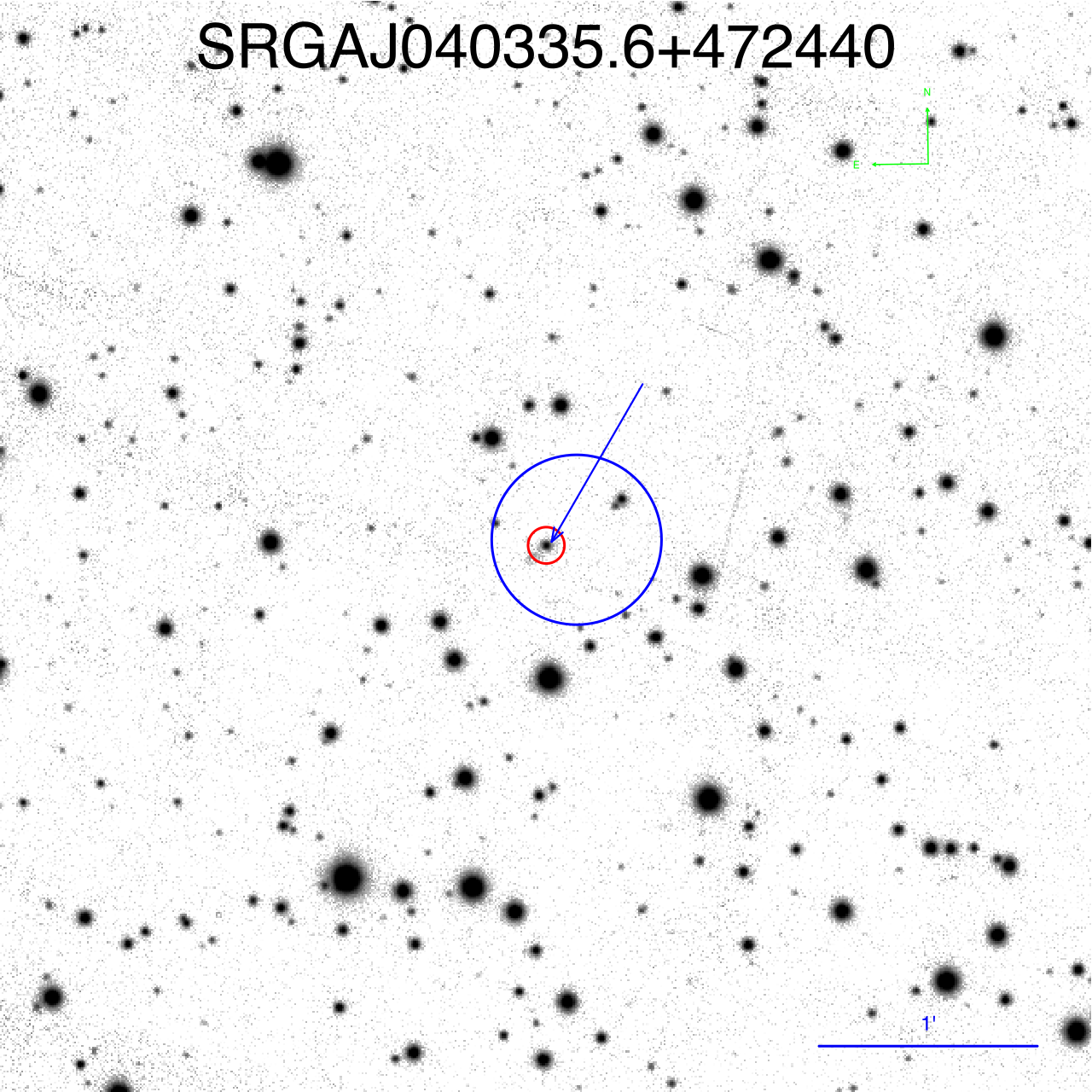}
  \includegraphics[width=0.25\columnwidth]{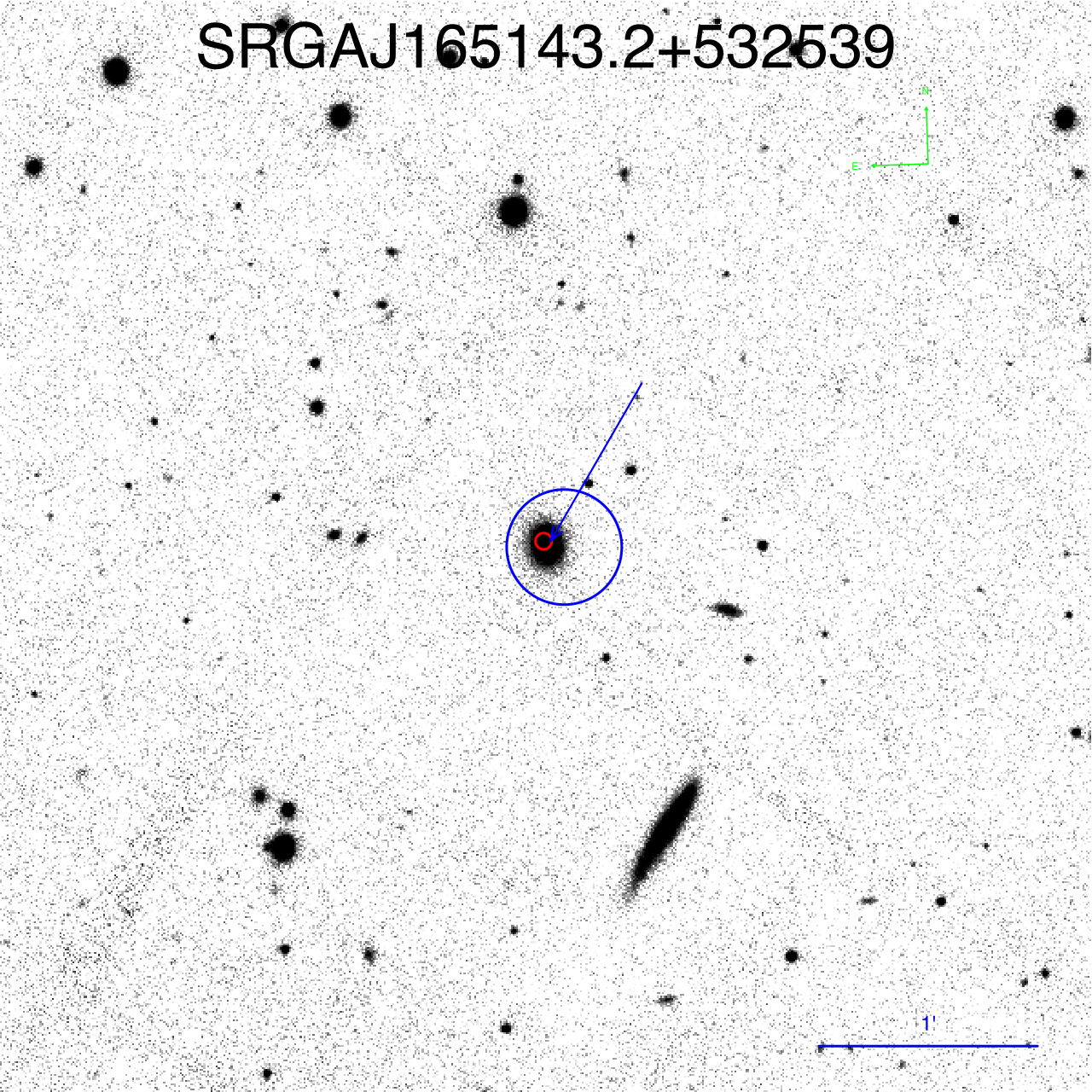}
  \includegraphics[width=0.25\columnwidth]{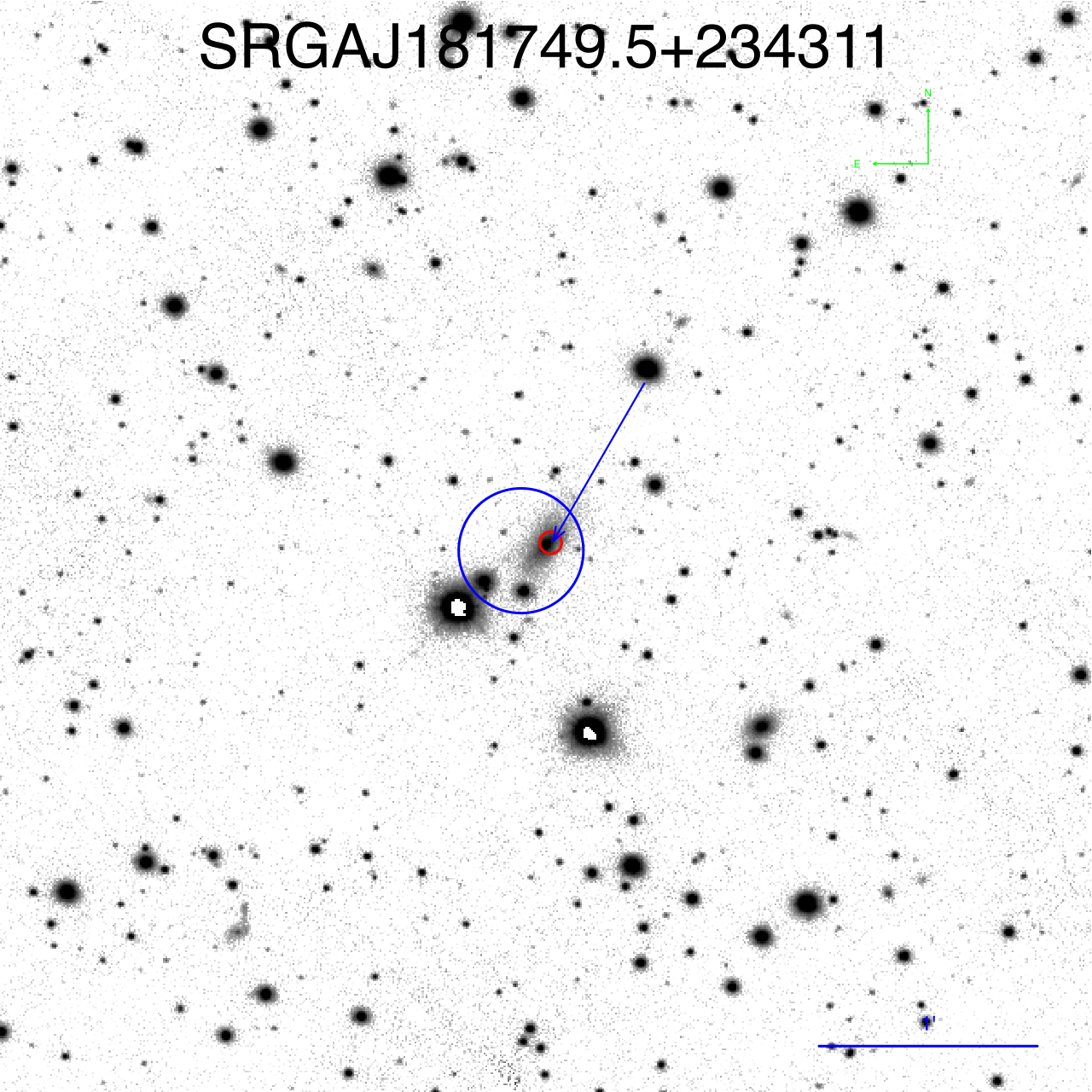}
  \includegraphics[width=0.25\columnwidth]{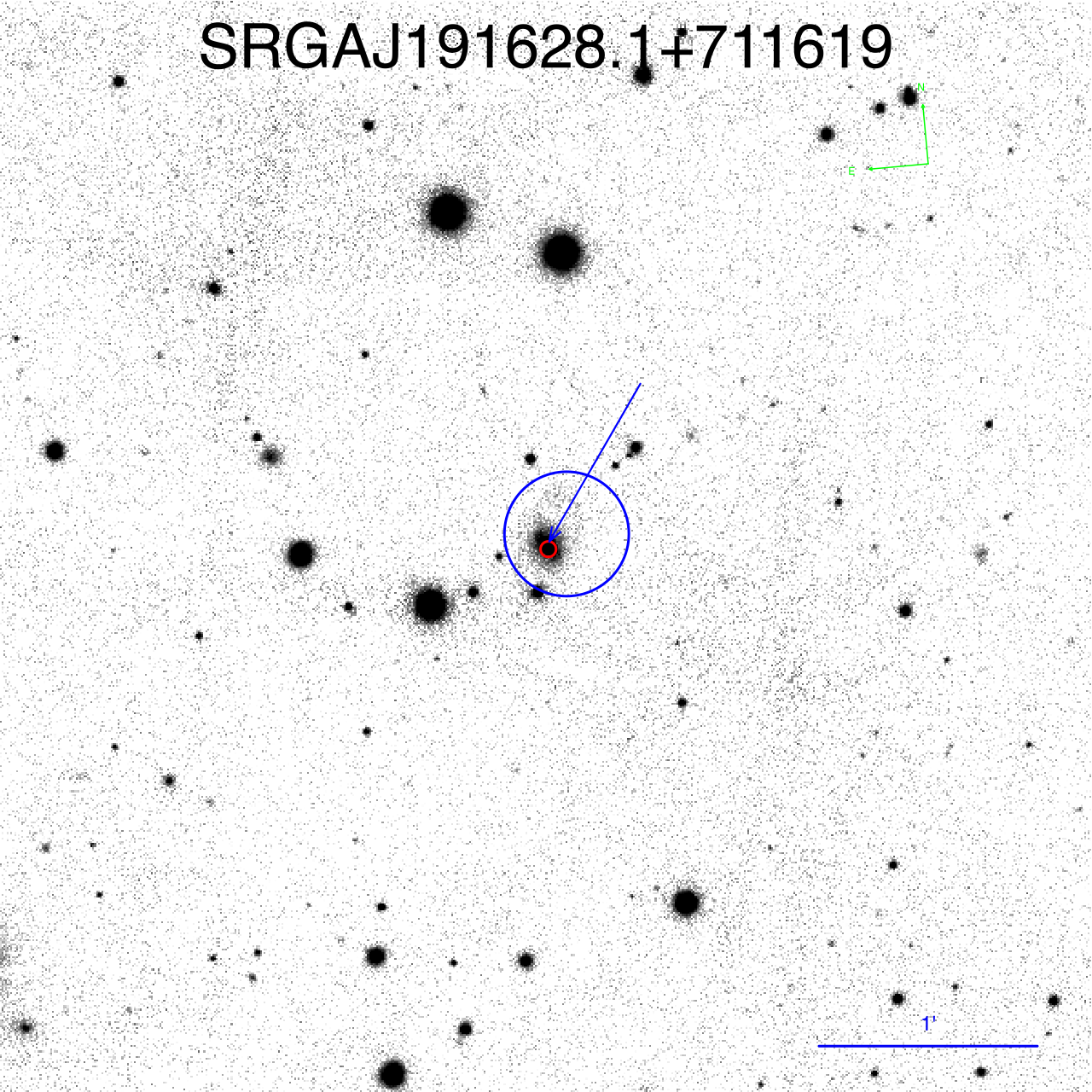}
  \includegraphics[width=0.25\columnwidth]{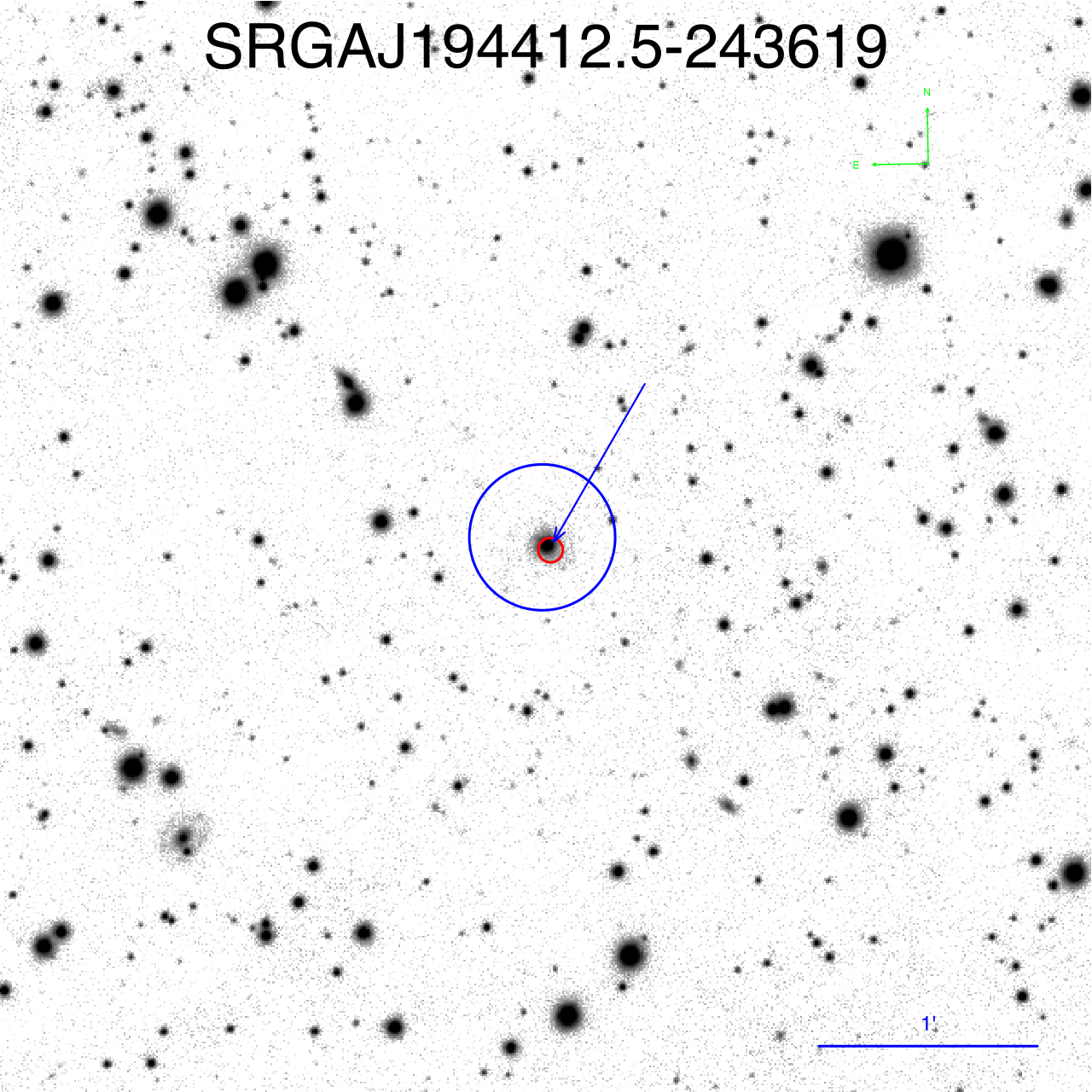}
  \includegraphics[width=0.25\columnwidth]{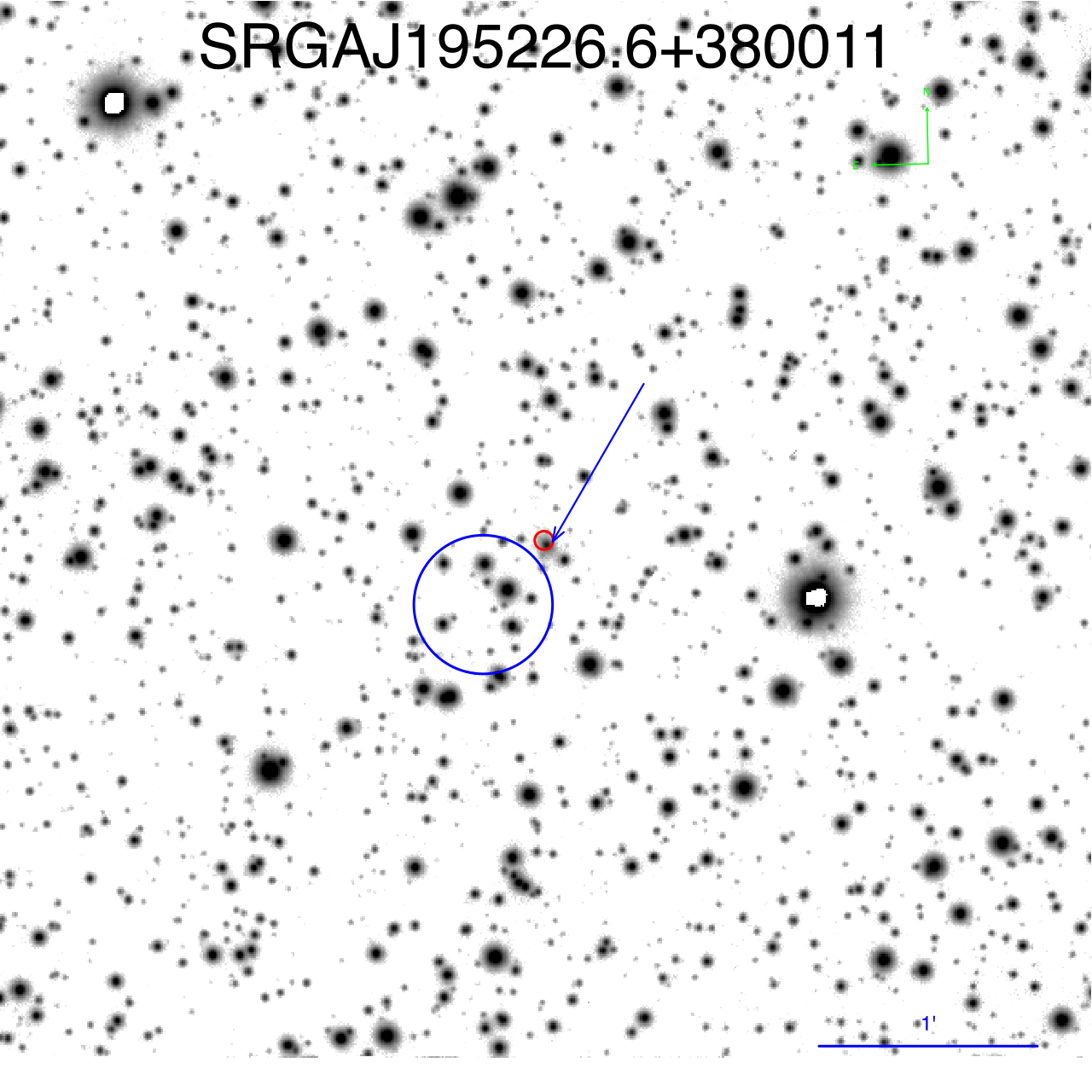}
  \includegraphics[width=0.25\columnwidth]{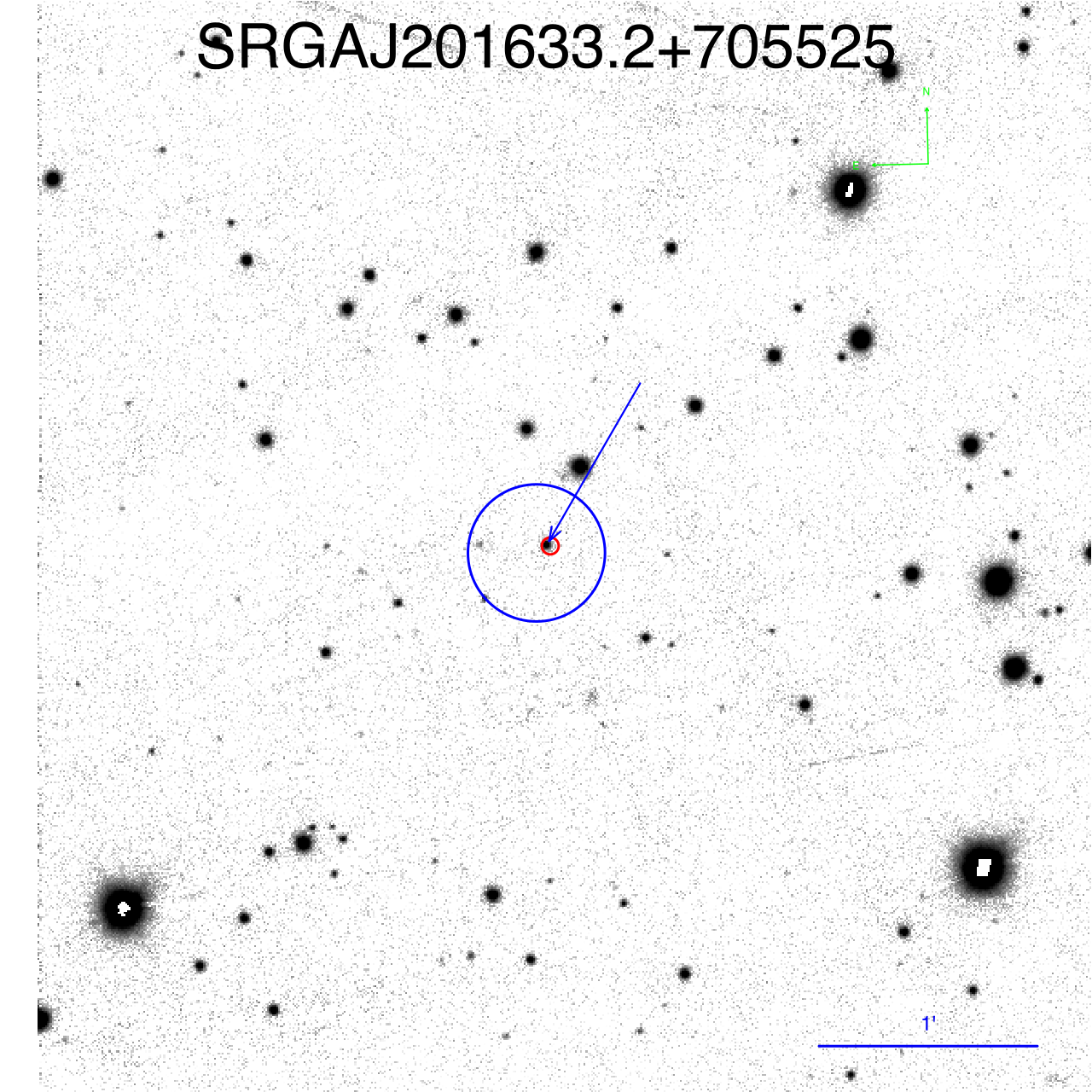}
  \caption{
  Optical images in the {\it r} filter from the PanSTARRS PS1 survey \citep{chambers2016}. The large and small circles indicate the \art\ and \ero\ position error circles of the X-ray sources, respectively (see $\ra$ and $\re$ in Table~\ref{tab:sources}). The arrows indicate the optical objects whose spectra are analyzed in this paper. The horizontal line corresponds to 1$\arcmin$.
  }
  \label{fig:guid_images}
\end{figure*}

Figure~\ref{fig:guid_images} shows optical images of the objects with the overlaid \art\ and \ero\ position error circles of the X-ray sources. A specific optical counterpart can be unambiguously associated with each X-ray source. All these objects were classified in the \ass\ catalog as AGN candidates based on the available (mostly photometric) archival data, such as the extent of the optical image, the characteristic infrared (IR) color, and the presence of radio emission. Note that in the case of SRGA\,J195226.6+380011 the optical counterpart is 4.8\arcsec\ outside the radius of the \art\ position error circle $\ra=19.0$. This is probably related to the systematic error of the astrometry in the \ass\ catalog, which is estimated to be $\sim 7\arcsec$ \citep{sazonov2024}.

Information about the properties of the objects being studied in the optical, IR, and radio bands, namely the coordinates of the optical counterpart, the IR color $W1-W2$ from the data of the ALLWISE catalog \citep{allwise}, and the radio flux at 1.4\,GHz or in the band 2--4~GHz, if the source is present in the NVSS \citep{nvss} or VLASS \citep{vlass} catalog, respectively, is collected in Table~\ref{tab:multiwave}. For all of the objects except SRGA\,J023800.1+193818 $W1-W2>0.5$. This points to the probable activity of the galactic nucleus associated with the accretion of matter onto a supermassive black hole \citep{stern12}. The IR color of SRGA\,J023800.1+193818 atypical of AGNs is apparently related to the low luminosity of the active nucleus in this galaxy, as discussed below in the Section ``Comments on Individual Objects''. Radio emission is detected from five objects, with SRGA\,J000132.9+240237 being a powerful radio source.

\section{OPTICAL OBSERVATIONS}
\label{s:optical}

\begin{table*}
\caption{Log of optical observations}
\label{tab:opt_obs}
\centering
\renewcommand{\arraystretch}{1.5}
\begin{tabular}{llccl}
\toprule
№ & \art\ source           & Date       & Grism   & Exposure time, s \\
\midrule
1 & SRGA\,J000132.9$+$240237$^*$ &  2013-09-08 & Channel B & $4\times3600 $ \\
  &                              &  2013-09-08 & Channel R & $4\times3600 $ \\
2 & SRGA\,J001059.5$+$424341 & 2023-11-09 & VPHG600G & $4\times600$ \\
  &                          & 2023-11-09 & VPHG600R & $4\times600$ \\
3 & SRGA\,J023800.1$+$193818 & 2023-11-08 & VPHG600G & $3\times600$ \\
4 & SRGA\,J025900.3$+$502958 & 2023-09-19 & VPHG600G & $4\times300$ \\
  &                          & 2023-09-19 & VPHG600R & $3\times300$ \\
5 & SRGA\,J040335.6$+$472440 & 2023-11-13 & VPHG600G & $4\times600$ \\ 
6 & SRGA\,J165143.2$+$532539 & 2023-04-13 & VPHG600G & $4\times300$ \\
7 & SRGA\,J181749.5$+$234311 & 2023-05-19 & VPHG600G & $4\times300$ \\
8 & SRGA\,J191628.1$+$711619 & 2023-04-13 & VPHG600G & $4\times600$ \\
9 & SRGA\,J194412.5$-$243619$^+$ & 2003-08-04 & V & $6\times1200$ \\
  &                          & 2003-08-04 & R & $5\times600$ \\
10 & SRGA\,J195226.6$+$380011 & 2023-11-09 & VPHG600R & $4\times600$ \\
11 & SRGA\,J201633.2$+$705525 & 2023-11-09 & VPHG600G & $5\times600$ \\

\bottomrule
\end{tabular}
\begin{flushleft}
  $^*$ -- SDSS, $^+$ -- 6dF.
\end{flushleft}
\end{table*}

Our optical observations of the objects were carried out in 2023 with the 1.6-m \azt\ telescope at the Sayan Observatory (the Institute of Solar–Terrestrial Physics, the Siberian Branch of the Russian Academy of Sciences) using the low- and medium-resolution ADAM spectrograph \citep{adam16,adam16a}. Table~\ref{tab:opt_obs} presents a log of observations.

Our observations were carried out using a slit with a width of 2$\arcsec$ and a position angle of $0^{\circ}$. As the spectrograph’s dispersive element we used two volume phase holographic gratings (grisms): VPHG600G with the spectral range 3800\AA\,--\,7250\AA\ and a resolution $R \approx 900$ and VPHG600R with the spectral range 6450\AA\,--\,10000\AA\ and a resolution $R \approx 1300$ For each source we first performed the observations with the VPHG600G grism, and if the wavelength of the $H_{\alpha}$ emission line turned out to be outside the spectral range of VPHG600G after the spectroscopic imaging or after the processing of the spectra, then we also performed the observations with the VPHG600R grism. For the observations of SRGA\,J195226.6$+$380011 we used only the VPHG600R grism.

The observations were carried out at a seeing no poorer than 2.5$\arcsec$. The spectrophotometric standards were observed at dusk and dawn on each observing night. We used the list of standards from the site of the European Southern Observatory\footnote{https://www.eso.org/sci/observing/tools/standards/ \newline /spectra/stanlis.html}. After the series of spectroscopic images for each object, we took two or three images of calibration lamps with a line and continuum spectrum. The data were processed with the \textit{PyRAF} package \footnote{https://iraf-community.github.io/pyraf.html} and our own software. The spectra of each object were corrected for interstellar extinction \citep{extinct_law}. The color excess $E(B-V)$ was calculated with the help of the software at the \textit{GALExtin} site\footnote{http://www.galextin.org/}. We used the dust reddening map from \cite{Schlafly14} and the coefficient $R_v = 2.742$ from \cite{Schlafly11}.

For two objects, SRGA\,J000132.9+240237 and SRGA\,J194412.5$-$243619, we performed no optical observations, since the archival SDSS \citep{sdssdr16} and 6dF \citep{6dftotal} spectroscopic data, respectively, were already available for them. We carried out the spectral classification based on these data. SDSS was conducted with the 2.5-m wide-field telescope at the Apache Point Observatory using two optical-fiber BOSS spectrographs with two cameras with overlapping spectral ranges, 3600--6350~\AA\ (B) and 5650--10000~\AA\ (R). The 6dF survey was conducted at the UKST 1.2-m Schmidt telescope using a multifiber spectrograph with a 5.7$^\circ$ field of view equipped with two low-resolution ($R\approx 1000$) gratings with overlapping spectral ranges (V and R). The range 4000--7500~\AA was completely covered. The spectra from the 6dF survey were not flux-calibrated and are presented in counts, which does not allow the absolute fluxes in emission lines to be measured. However, these data can be used to estimate the line equivalent widths and the ratios of the fluxes in pairs of closely spaced lines, which is enough for the classification of AGNs.

\section{X-Ray Data}
\label{s:xray}

Depending on their positions in the sky, the sources being studied were scanned by the \ero\ and \art\ telescopes four or five times (3 and 7 sources, respectively) with an interval of half a year. The source SRGA\,J040335.6+472440 was observed five times by \art, but during the fifth scan (February 27–March 5, 2022) \ero\ was switched off. Based on the set of data from all scans, we constructed the spectra of the sources in a wide energy range, 0.2--20~keV.

The \ero\ data were processed with the calibration and data processing system created and maintained at the Space Research Institute of the Russian Academy of Sciences, which uses the elements of the eSASS (eROSITA Science Analysis Software System) package and the software developed by the science group on the X-ray catalog of the Russian \ero\ consortium. We extracted the source spectra in a circle of radius 60$\arcsec$ and the background spectra in a ring with an inner radius of 120$\arcsec$ and an outer radius of 300$\arcsec$ around the source. If other sources fell into the background region, then the photons in a region of radius 40$\arcsec$ around them were excluded. The spectra were extracted from the data of all seven ART-XC modules in the energy range 0.2--9.0~keV. When fitting the spectra, the data were binned in such a way that there were at least 3 counts in each energy channel.

To construct the spectra of the sources from the \art\ data, we used the counts collected in three broad energy bands, 4--7, 7--12 and 12--20, from a region of radius 71$\arcsec$ (in which 90\% of the photons from the source during the sky survey are concentrated). The data from all seven \art\ modules were combined. In our spectral analysis we used the diagonal response matrix constructed from Crab Nebula observations. The background level was estimated using the data in the hard 30--70~keV energy band and the survey image wavelet decomposition maps (see \citealt{sazonov2024}).

\section{Results}
\label{s:results}

\subsection{Optical Spectra}

\begin{figure*}
  \centering
  
    №1: SRGA\,J000132.9$+$240237
  \includegraphics[width=0.8\columnwidth]{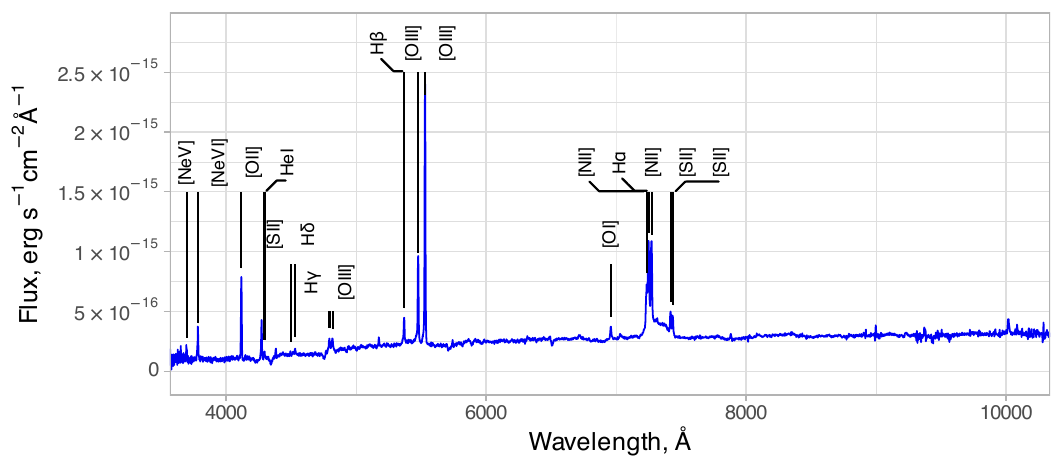}
  
    №2: SRGA\,J001059.5$+$424341
  \includegraphics[width=0.8\columnwidth]{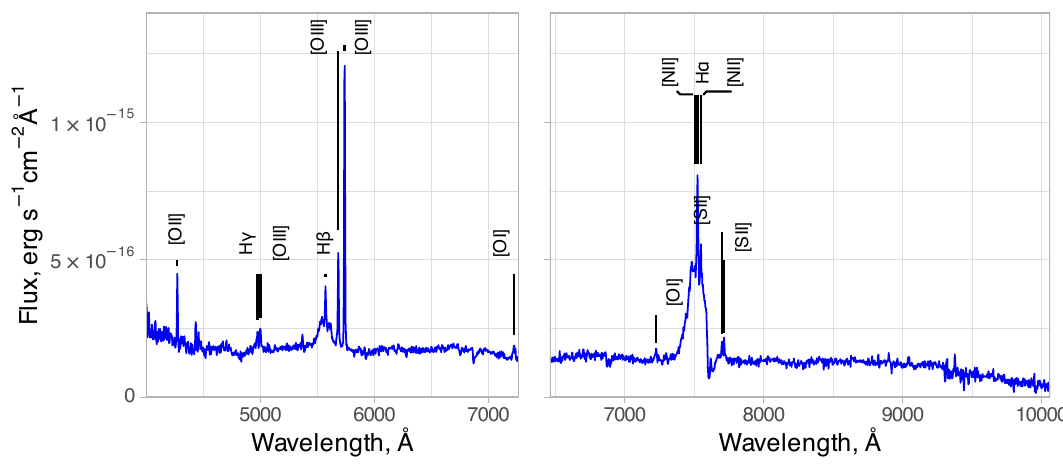}
  
    №3: SRGA\,J023800.1$+$193818
  \includegraphics[width=0.8\columnwidth]{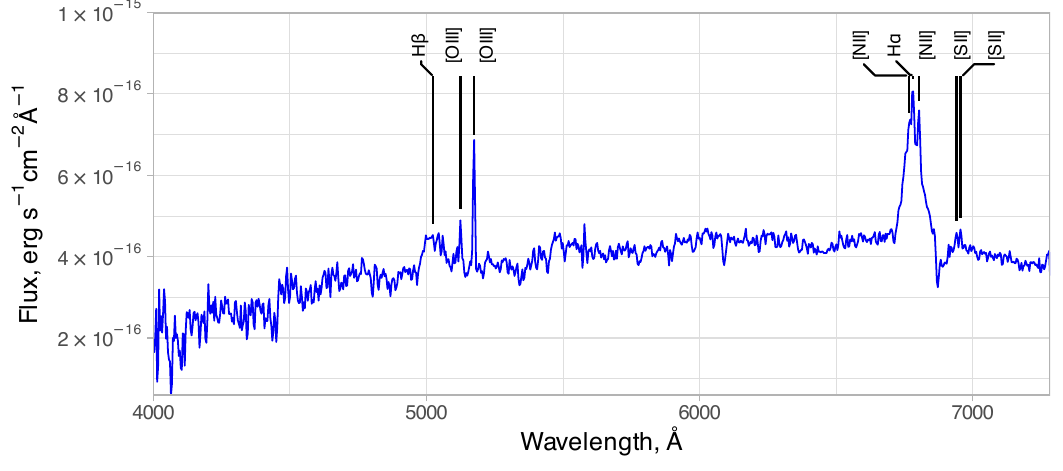}
  \caption{
  Optical spectra with labeled main emission and absorption lines.
  }
  \label{fig:spec0001_0238}
\end{figure*}
\addtocounter{figure}{-1}
\begin{figure*}
  \centering
  \vfill
  
  №4: SRGA\,J025900.3$+$502958
  \includegraphics[width=0.8\columnwidth]{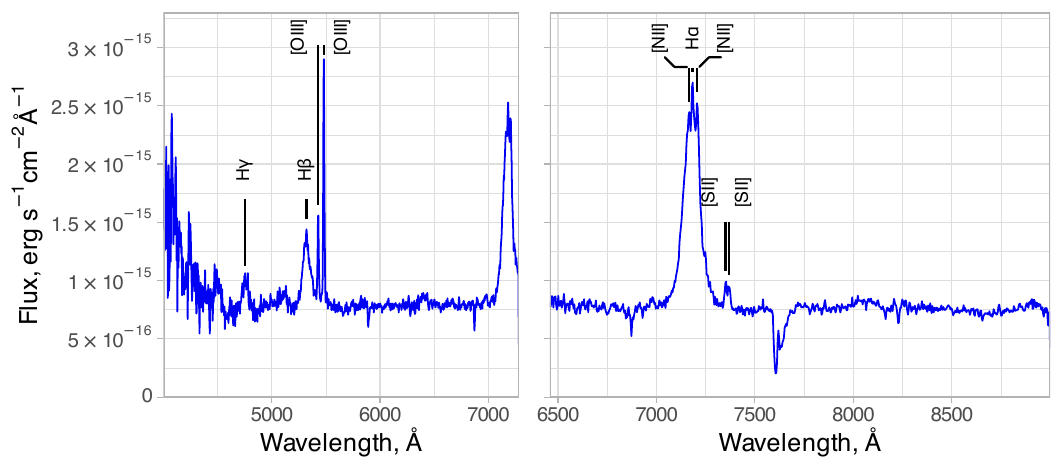}
  
  №5: SRGA\,J040335.6$+$472440
  \includegraphics[width=0.8\columnwidth]{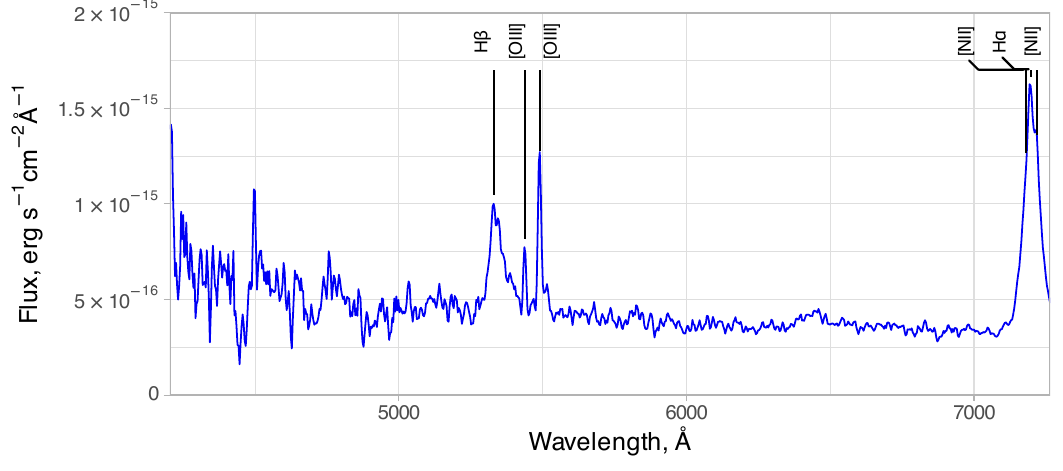}
  
  №6: SRGA\,J165143.2$+$532539
  \includegraphics[width=0.8\columnwidth]{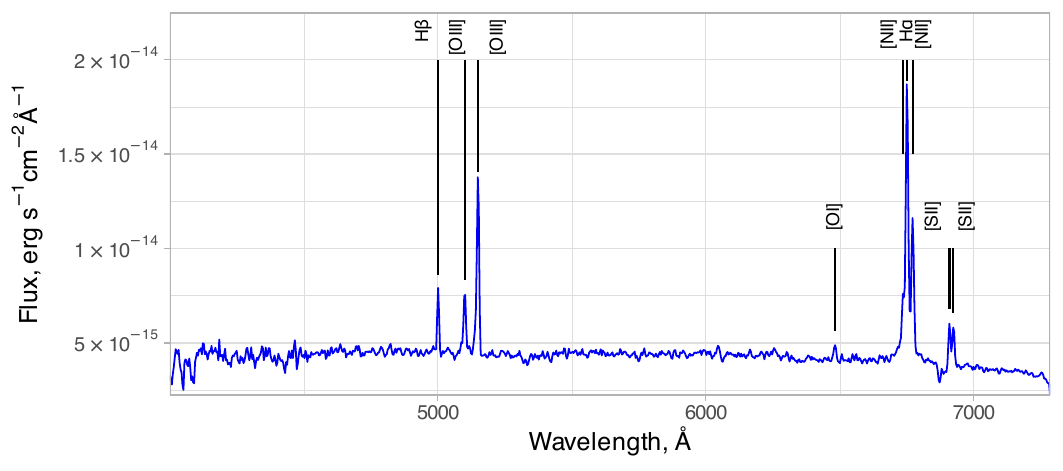}
  \caption{
  (Contd.)
  }
  \label{fig:spec0259_1640}
\end{figure*}
\addtocounter{figure}{-1}
\begin{figure*}
  \centering
  \vfill
  
  №7: SRGA\,J181749.5$+$234311
  \includegraphics[width=0.8\columnwidth]{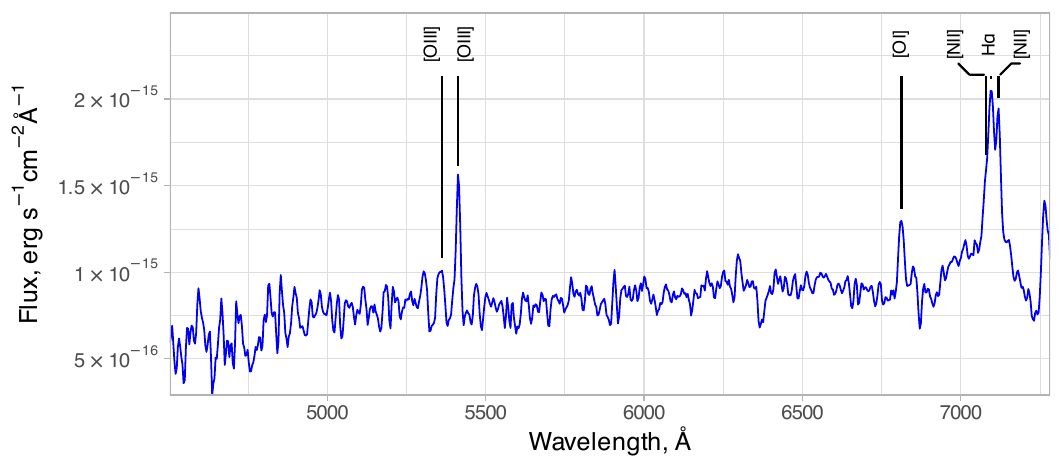}
  
  №8: SRGA\,J191628.1$+$711619
  \includegraphics[width=0.8\columnwidth]{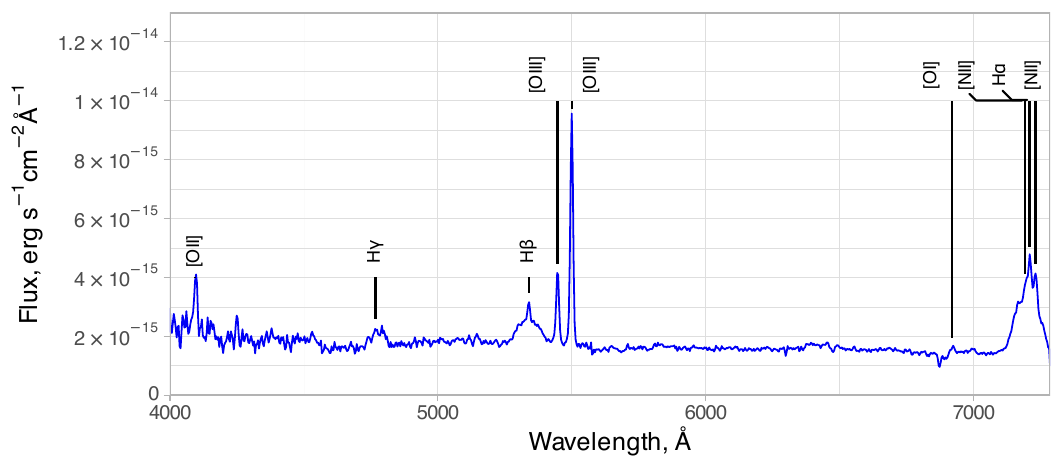}
  
  №9: SRGA\,J194412.5$-$243619
  \includegraphics[width=0.8\columnwidth]{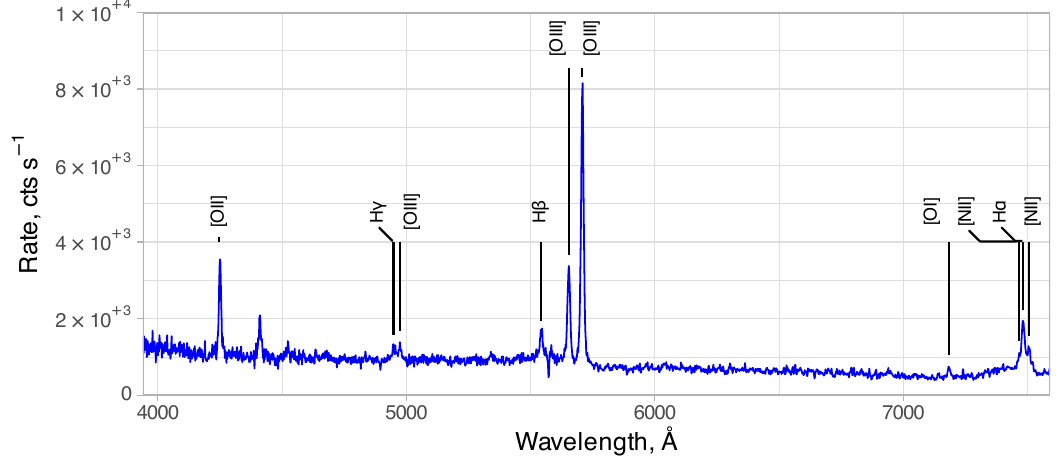}
  \caption{
  (Contd.)
  }
  \label{fig:spec1807_1838}
\end{figure*}
\addtocounter{figure}{-1}
\begin{figure*}
  \centering
  \vfill
  №10: SRGA\,J195226.6$+$380011
  \includegraphics[width=0.8\columnwidth]{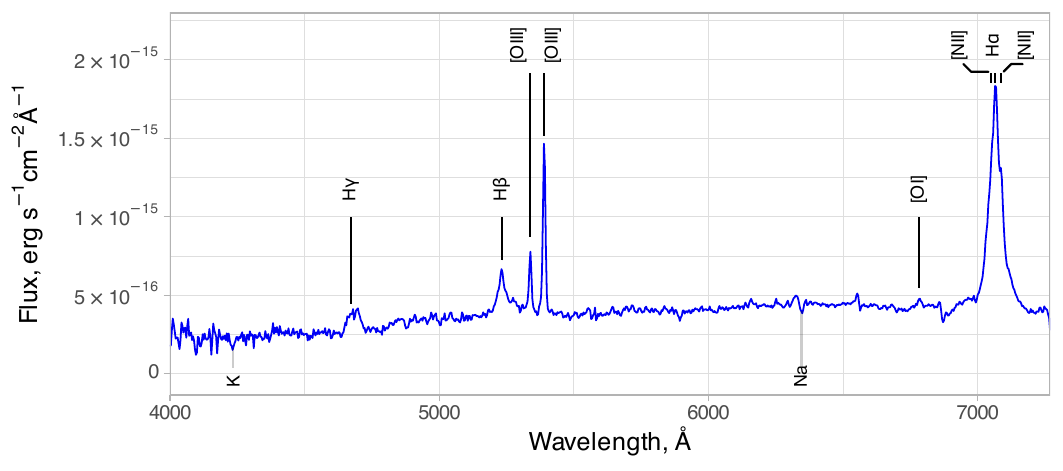}
  
  №11: SRGA\,J201633.2$+$705525
  \includegraphics[width=0.8\columnwidth]{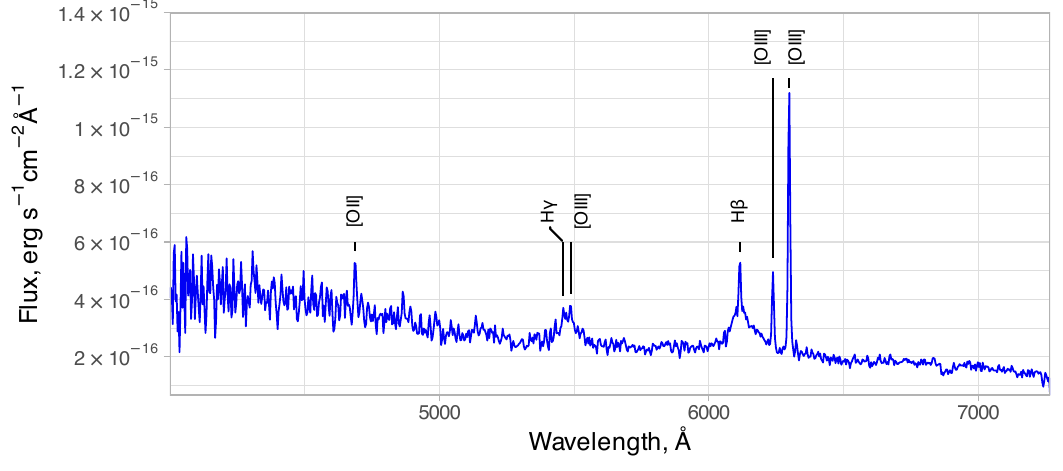}
  \caption{
  (Contd.)
  }
  \label{fig:spec1916_1952}
\end{figure*}

Figure~\ref{fig:spec0001_0238} shows optical spectra of the objects being studied. Standard criteria based on the emission line flux ratios \citep{oster, veron} were used to classify the Seyfert galaxies. We fitted the spectral continuum by a polynomial and the emission lines by Gaussians. Thus, for each line we determined the central wavelength, the full width at half maximum $\fwhmmes$, the flux, and the equivalent width $EW$. The FWHM of the broad Balmer lines was corrected for the spectral resolution of the instrument: $\fwhm=\sqrt{\fwhmmes^2-\fwhmres^2}$, where $\fwhmres$ was determined for each dispersive element and each slit as the FWHM of the lines in the calibration lamp spectrum

Table~\ref{tab:lines} presents the measured line characteristics for all sources. The errors are given at 68\% confidence. The confidence interval for the redshift was determined as the error of the mean narrow-line redshift. The measured FWHM of the narrow emission lines are consistent with the instrumental broadening, and, therefore, the values of FWHM are not given for them. The confidence intervals for the line equivalent widths ($EW$) were obtained by the Monte Carlo method. Assuming that the flux errors obeyed a normal distribution, we selected 1000 spectrum realizations. Then, for each of the realizations we estimated $EW$. The confidence intervals were estimated from the derived EW distribution. To obtain an upper limit on the line flux, we fixed the center of the Gaussian and took its width to be equal to the instrumental broadening.

The results of our classification of the sources and the measured redshifts are presented in Table~\ref{tab:class}.

\subsection{X-ray Spectra}

\begin{figure*}
  \centering
  \includegraphics[width=0.4\columnwidth]{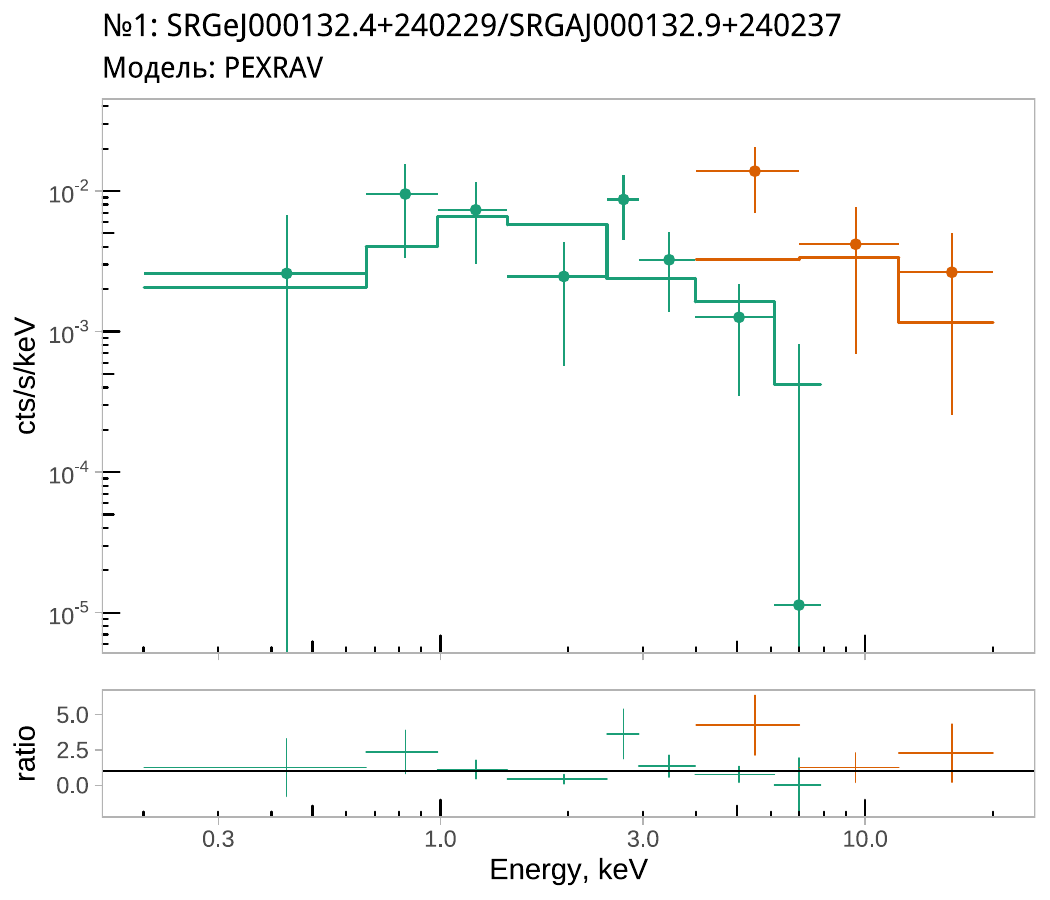}
  \includegraphics[width=0.4\columnwidth]{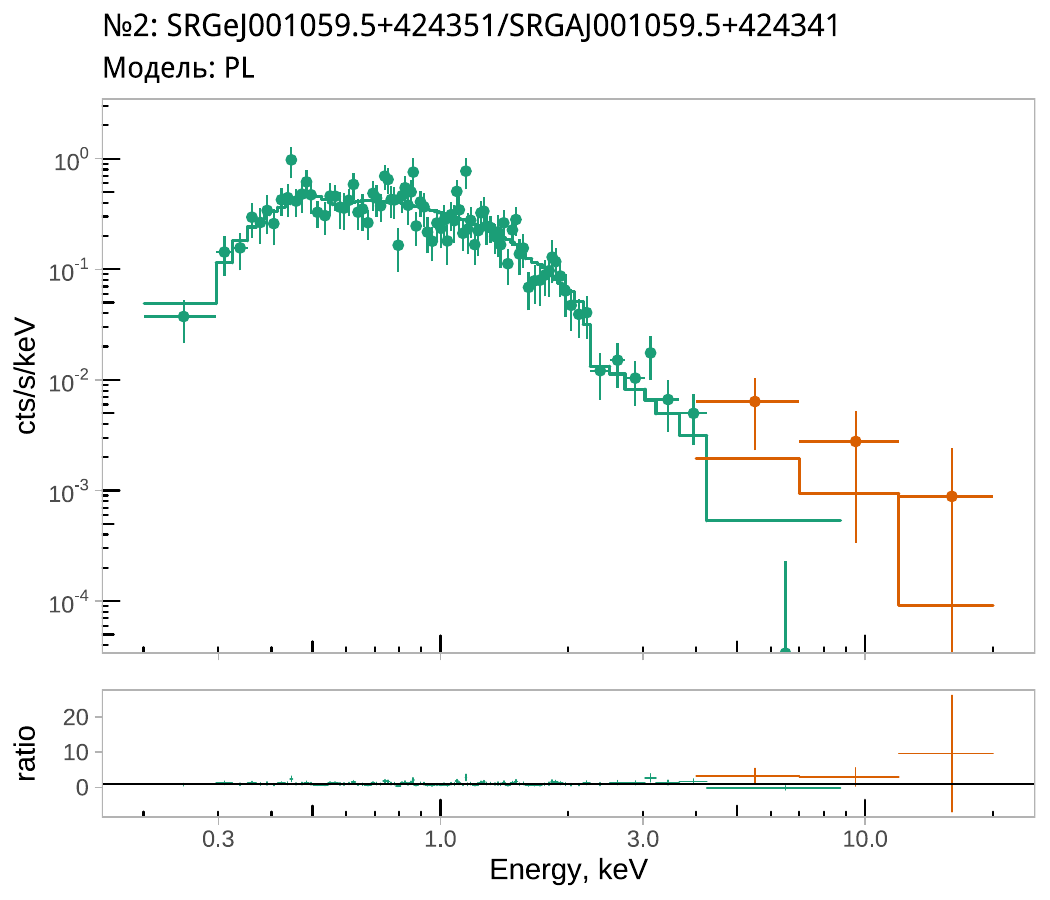}
  \includegraphics[width=0.4\columnwidth]{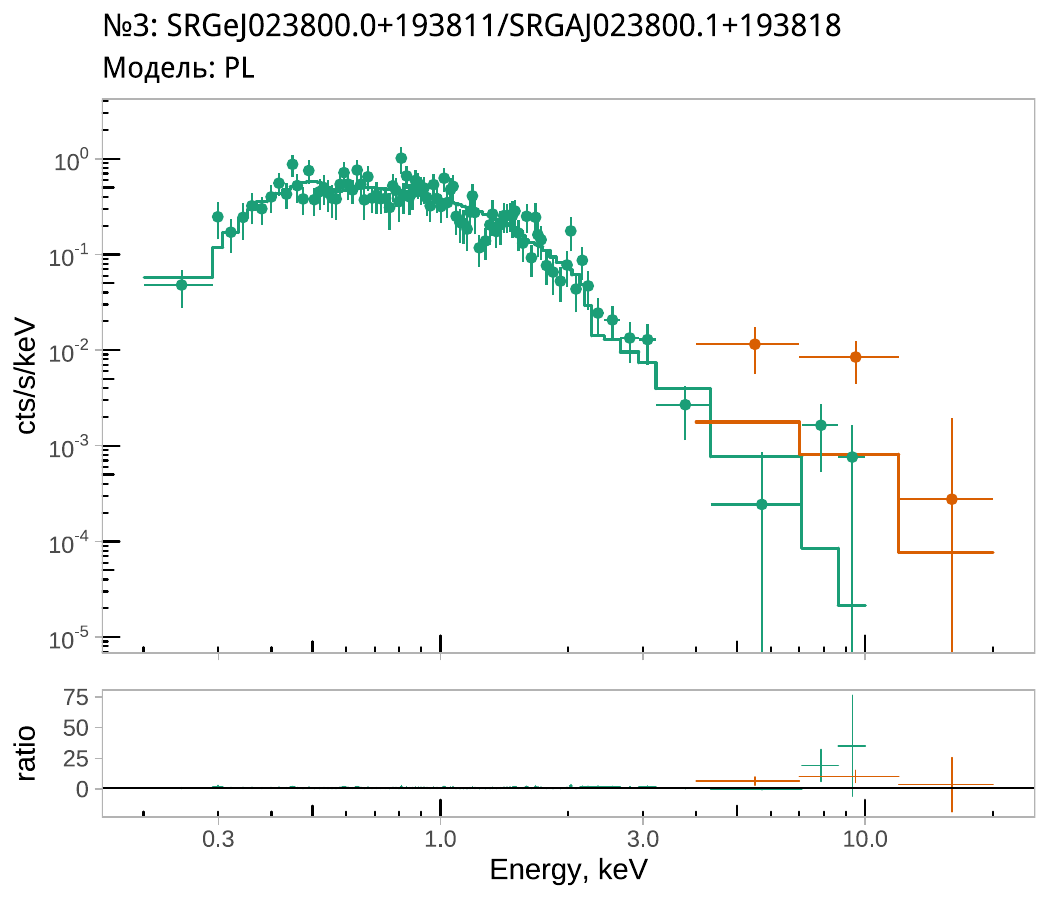}
  \includegraphics[width=0.4\columnwidth]{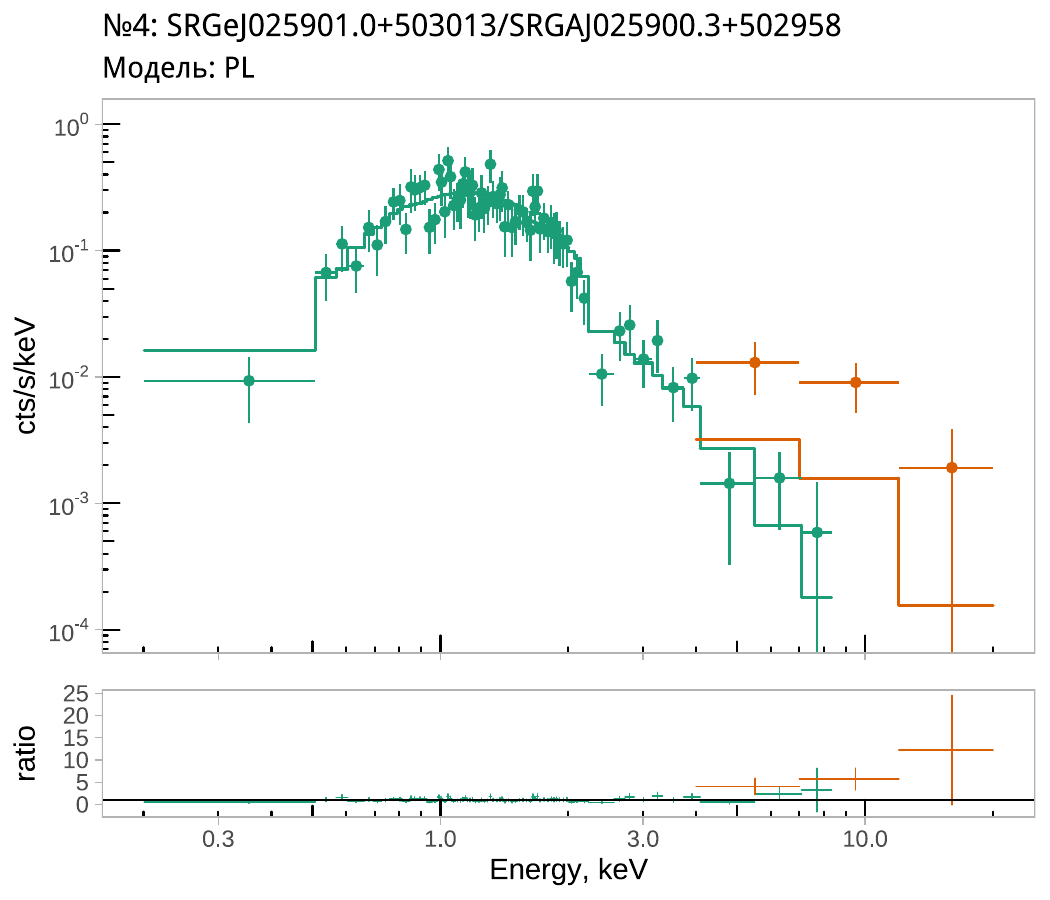}
  \includegraphics[width=0.4\columnwidth]{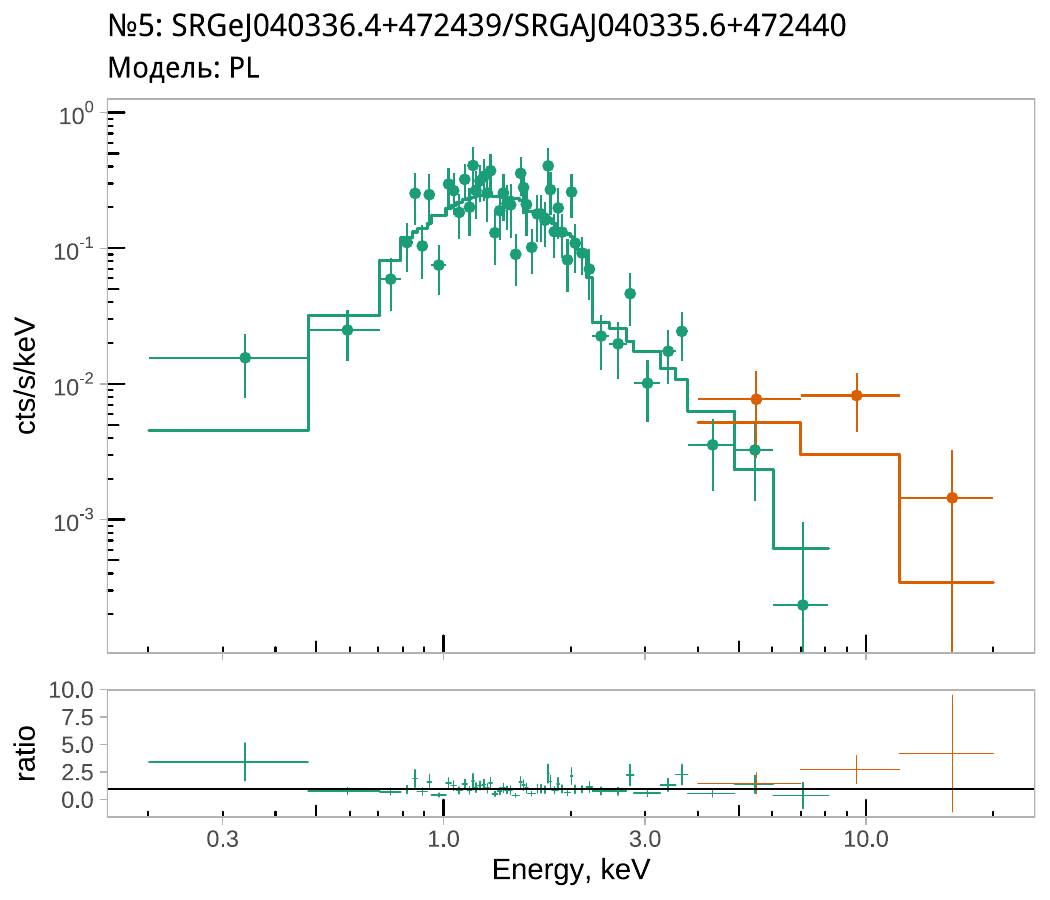}
  \includegraphics[width=0.4\columnwidth]{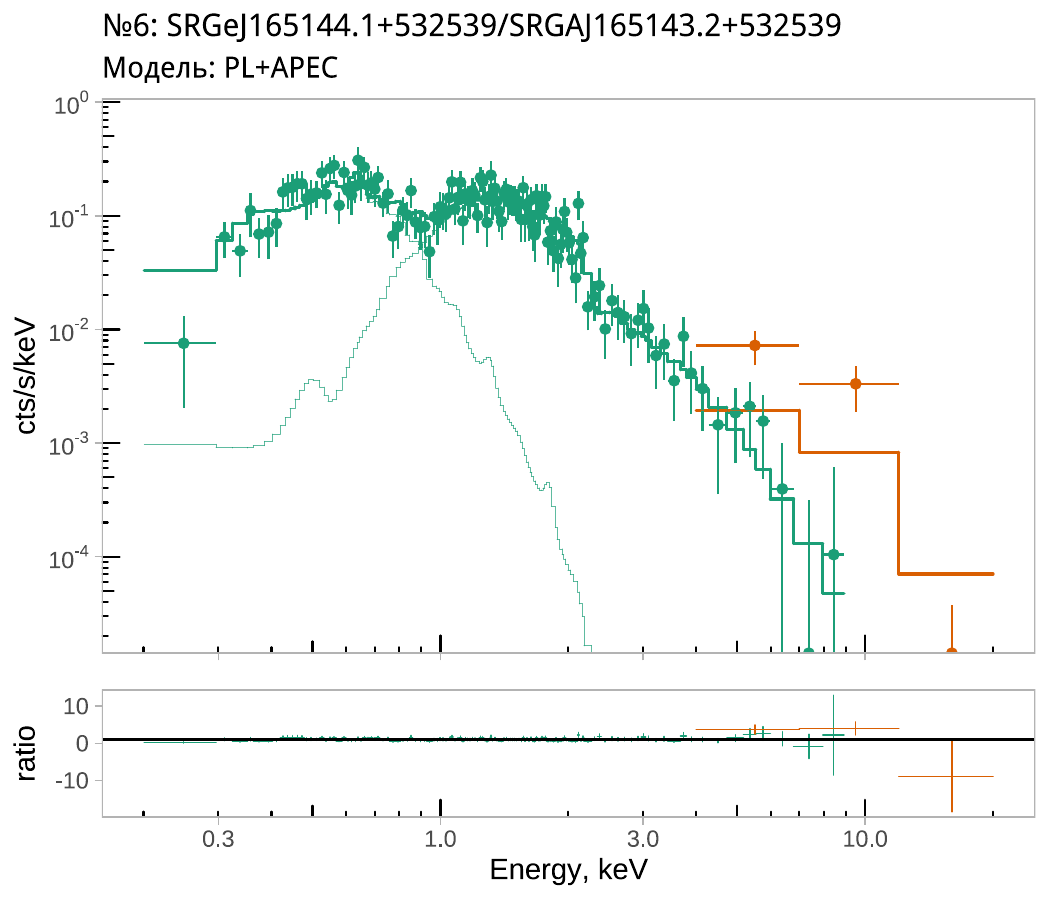}
  \caption{
   X-ray spectra from the\ero\ (green) and \art\ (orange) data and the best-fit models from the \ero\ data (see Table~\ref{tab:xray_params}). For source no. 6 we used the two-component {\sc PL+APEC} model; therefore, in addition to the full model, both components are also shown. The ratio of the measurements to the model is shown on the lower panels.
  }
  \label{fig:xray_plots}
\end{figure*}

\addtocounter{figure}{-1}

\begin{figure*}
  \centering
  \includegraphics[width=0.4\columnwidth]{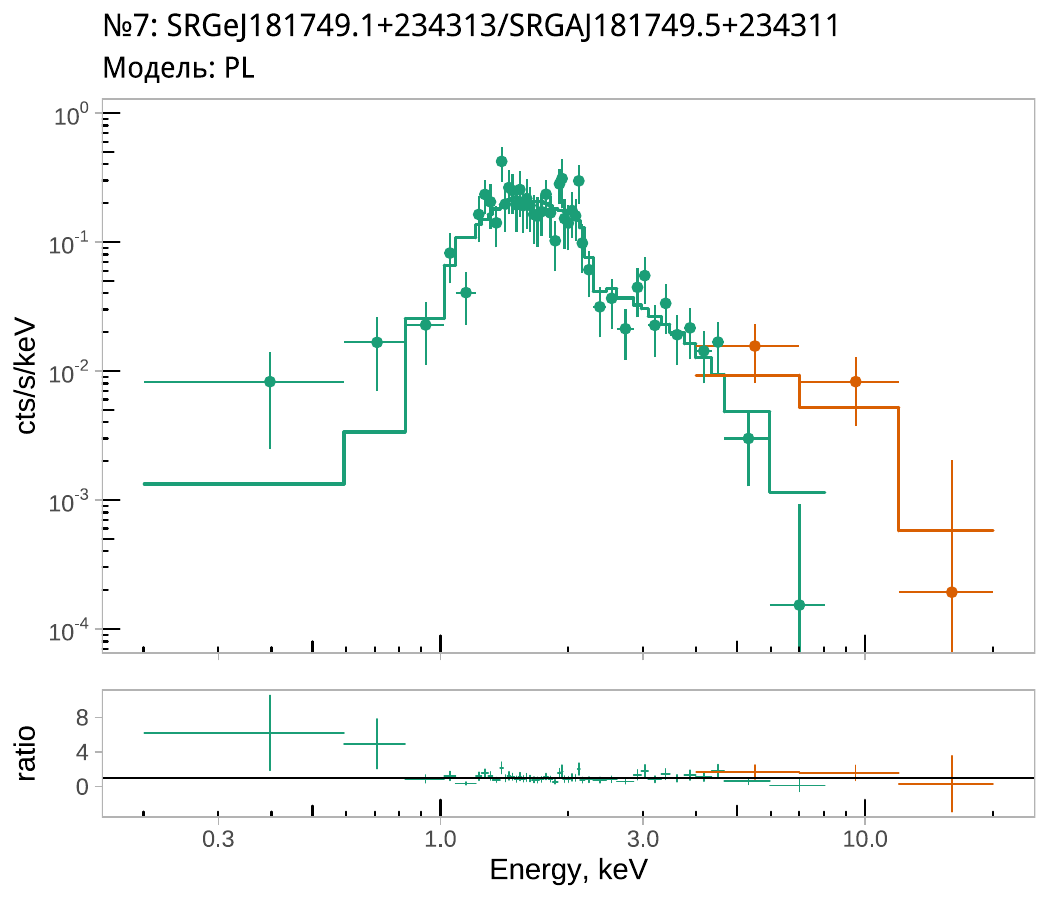}
  \includegraphics[width=0.4\columnwidth]{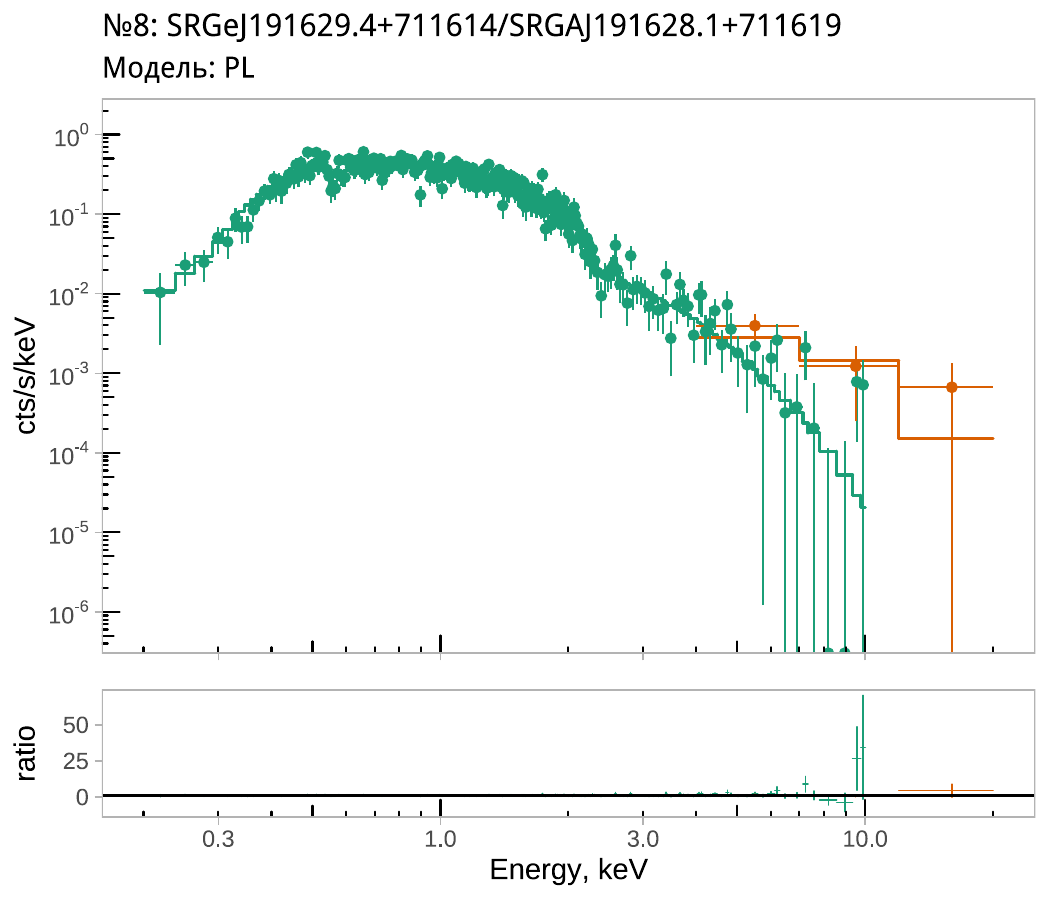}
  \includegraphics[width=0.4\columnwidth]{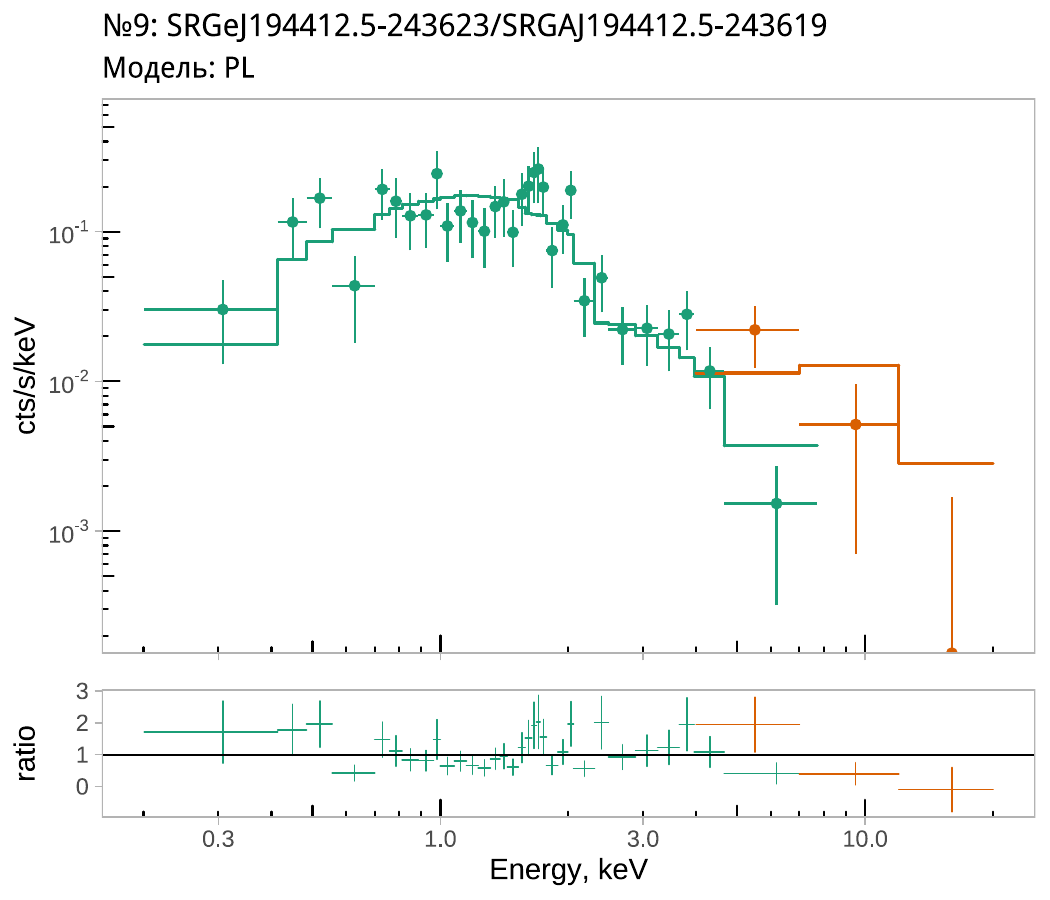}
  \includegraphics[width=0.4\columnwidth]{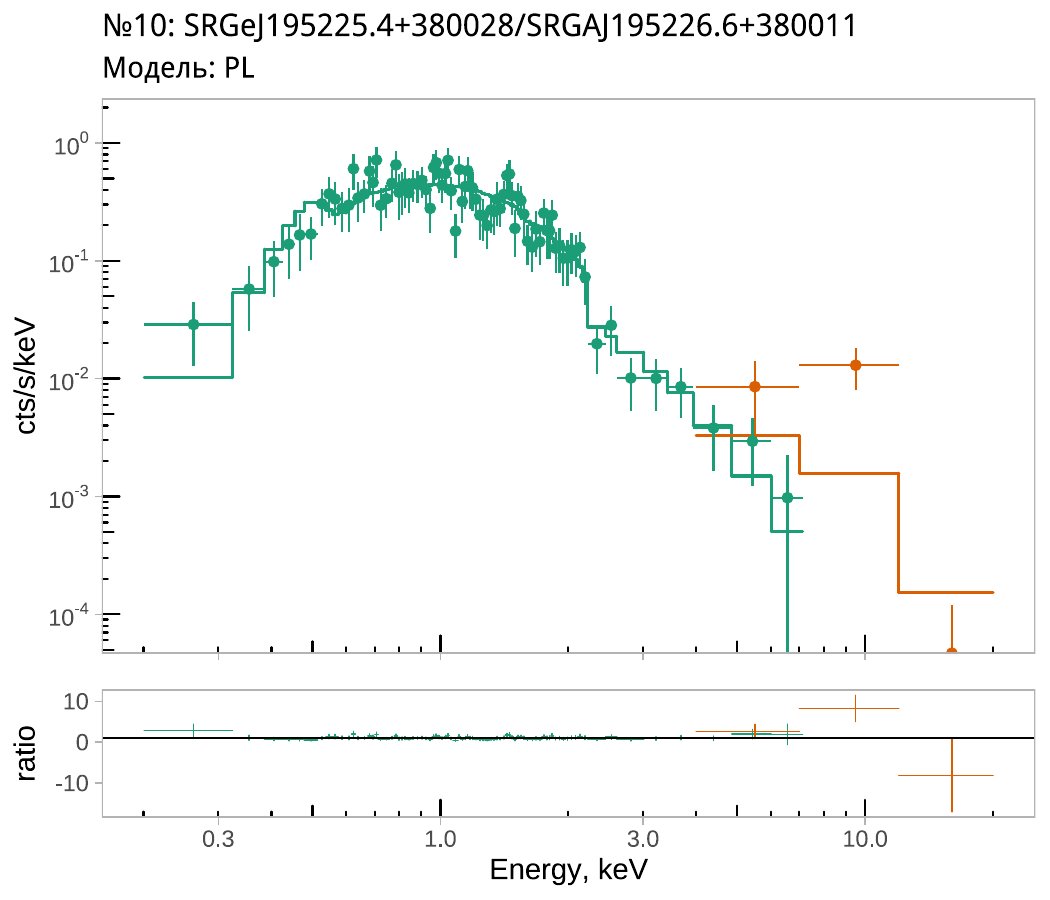}
  \includegraphics[width=0.4\columnwidth]{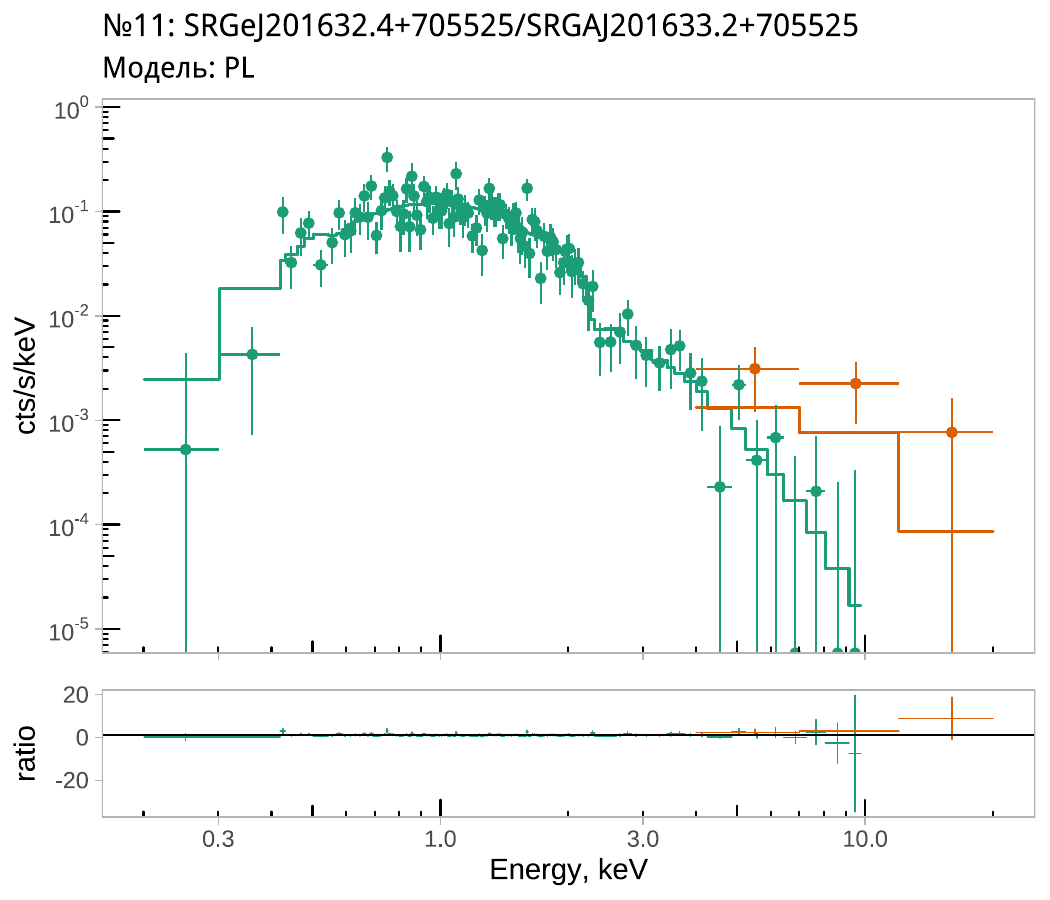}
  \caption{
  (Contd.)
  }
\end{figure*}

\begin{table*}
\caption{X-ray spectral parameters}
\label{tab:xray_params}
\flushleft
\renewcommand{\arraystretch}{2}
\renewcommand{\tabcolsep}{0.08cm}
\begin{tabular}{rccccccccrl}
\hline
\toprule
№ & Model &  $N_{\rm H}^{MW}$ & $N_{\rm H}$ & $F_{\rm 0.5-10}^{PL}$ & $\Gamma$ & $F_{\rm 0.5-10}^{PEXRAV}$    &  $cosI$ & $A$ & Cstat (dof)  \\
\hline
1 & PL & $0.04$ & $< 0.9$ & $14_{-10}^{+20}$ & $-0.5_{-0.8}^{+0.9}$ & & & & 11.3 (12)\\
  & PEXRAV & $0.04$ & & & $1.8$ (fix) & $10_{-4}^{+5}$ & $0.5$ (fix) & $1$ (fix) & 10.9 (14) \\
\midrule
\end{tabular}
\begin{tabular}{rcccccccccr}
№ & Model & $N_{\rm H}^{MW}$ & $N_{\rm H}$ & $F_{\rm 0.5-10}^{PL}$ & $\Gamma$ & $F_{\rm 0.5-10}^{APEC}$ & $kT$ & $A$ & Cstat (dof)\\
\hline
6 & PL & $0.05$ & $< 0.015$ & $19\pm2$ & $1.28_{-0.07}^{+0.12}$ & & & & 302.9 (256) \\
 & PL+APEC & $0.05$ & $0.9_{-0.2}^{+0.3}$ & $26_{-4}^{+9}$ & $2.5_{-0.4}^{+0.5}$ & $1.39\pm0.16$ & $0.24_{-0.02}^{+0.04}$ & $0.07_{-0.04}^{+0.06}$ & 248.3 (253) \\
\midrule
\end{tabular}
\begin{tabular}{rccccccr}
№ & Model & $N_{\rm H}^{MW}$ & $N_{\rm H}$ & $F_{\rm 0.5-10}^{PL}$ & $\Gamma$ & Cstat (dof)\\
\hline
2 & PL & $0.07$ & $< 0.05$ & $18.6_{-1.9}^{+2.1}$ & $2.20_{-0.15}^{+0.25}$ & 168.4 (171) \\
3 & PL & $0.08$ & $< 0.02$ & $18.6_{-1.7}^{+2.1}$ & $2.29_{-0.08}^{+0.18}$ & 168 (183) \\
4 & PL & $0.29$ & $0.11_{-0.08}^{+0.09}$ & $31\pm3$ & $2.2_{-0.3}^{+0.4}$ & 111.4 (140) \\
5 & PL & $0.52$ & $< 0.19$ & $39_{-5}^{+6}$ & $1.9_{-0.2}^{+0.4}$ & 110.3 (98) \\
7 & PL & $0.12$ & $1.6_{-0.4}^{+0.5}$ & $80_{-14}^{+30}$ & $2.0_{-0.5}^{+0.6}$ & 113.3 (88) \\
8 & PL & $0.09$ & $0.045_{-0.018}^{+0.019}$ & $24.1_{-1.1}^{+1.2}$ & $2.08\pm0.11$ & 358.4 (365) \\
9 & PL & $0.08$ & $< 0.05$ & $53_{-14}^{+12}$ & $0.66_{-0.17}^{+0.23}$ & 87.4 (61) \\
10 & PL & $0.22$ & $< 0.06$ & $32\pm3$ & $2.22_{-0.15}^{+0.31}$ & 138.3 (164) \\
11 & PL & $0.10$ & $0.16_{-0.06}^{+0.07}$ & $10.0_{-1.0}^{+1.4}$ & $1.9_{-0.2}^{+0.3}$ & 212.3 (215) \\
\bottomrule
\end{tabular}

$N_{\rm H}^{MW}$ and $N_{\rm H}$  are the gas column densities in the Galaxy and inside the AGN, respectively (in units of $10^{22}$~H atoms per cm$^2$); 
$F_{0.5-10}^{PL}$ is the 0.5–10 keV flux of the power-law component corrected for the Galactic and intrinsic absorption ($10^{-13}$~erg~s$^{-1}$~cm$^{-2}$); 
$\Gamma$ is the slope of the power-law component; 
$F_{0.5-10}^{PEXRAV}$ is the 0.5--10~keV flux of the \textsc{PEXRAV} component corrected for the Galactic absorption ($10^{-13}$~erg~s$^{-1}$~cm$^{-2}$); 
$cosI$ is the cosine of the inclination angle in the PEXRAV model; 
$A$ is the metal abundance with respect to the solar one; 
$F_{\rm 0.5-10}^{APEC}$ is the 0.5--10~keV flux of the {\sc APEC} component corrected for the Galactic absorption ($10^{-13}$~erg~s$^{-1}$~cm$^{-2}$); 
$kT$ is the plasma temperature in the APEC model (keV)
\end{table*}

We modeled the spectra based only on the \ero\ data, while no \art\ data were used. This is because the sources were selected by the detection significance in the \art\ all-sky survey, and the same survey data (i.e., only $\sim 10$ counts from each source) are considered here. The fluxes from the sources estimated from these data can be overestimated significantly due to the ``Eddington bias'' related to Poisson statistics (see, e.g., \citealt{wang04}). The \ero\ data are not subject to this effect, since they were not used when selecting the sources

The spectra were fitted in the 0.2--8~keV energy band with the XSPEC v12.12.0 software\footnote{https://heasarc.gsfc.nasa.gov/xanadu/xspec} \citep{arnaud96}. The $W$-statistic (using the $statistic~cstat$ option in XSPEC) that takes into account the X-ray background was used for our model fitting. To fit the spectra, we used several models described below.

Comptonization of the radiation from the accretion disk in its hot corona gives rise to power-law X-ray spectra with an exponential cutoff at high ($\gtrsim 100$~keV) energies \citep{sunyaev80,haardt91}. Therefore, we used the spectral model of a power-law continuum modified by photoabsorption in the Galaxy and inside the AGN (here-after the {\sc PL} model) as the basic one:

\begin{equation}
\textsc{tbabs} \times 
\textsc{ztbabs} \times 
\textsc{cflux} \times 
\textsc{zpowerlaw}.
\label{eq:pl}
\end{equation}
Here, {\sc tbabs} describes the absorption in the interstellar medium of the Galaxy toward the object being studied, and the neutral hydrogen column density from HI4PI \citep{bekhti16} is used; {\sc ztbabs} is responsible for the absorption inside the object itself (at its redshift); {\sc cflux} is the absorption-corrected flux in the observed 0.5--10~keV energy band.

The results of our modeling of the X-ray spectra for the sources are presented in Fig.~\ref{fig:xray_plots} and Table~\ref{tab:xray_params}. The 90\% confidence intervals for the parameters are given. The \ero\ data in the figures were rebinned into wider energy bins for clarity. Apart from the \ero\ measurements, the fluxes from the sources in the 4--7, 7--12 and 12--20~keV energy bands estimated from the \art\ data are also shown. The latter are located, on average, higher than the corresponding model spectra, which is obviously related to the above-mentioned Eddington bias.

\subsubsection{The soft X-ray excess in the spectrum of SRGA\,J165143.2$+$532539/SRGeJ165144.1+53253}

Noticeable residuals are observed at low energies when fitting the spectrum of source no. 6 (optical type Sy1.9) by the PL model, suggesting the presence of an additional soft component. In the X-ray spectra of type 2 AGNs additional radiation at energies below 2~keV is often observed (see, e.g., \citealt{guainazzi05,guainazzi07}) against the background of an absorbed power-law continuum. This can be both the radiation from the central source scattered and reprocessed (as a result of photoionization and recombination) in the gas outside the dusty torus and the intrinsic X-ray radiation from the stellar population and the interstellar medium of the host galaxy.

We attempted to describe the soft X-ray excess in the spectrum of source no. 6 using the APEC model designed to describe the thermal radiation spectra of a hot, optically thin plasma \citep{apec}. In the language of XSPEC the corresponding two-component model is written as (here after the PL+APEC model):
\begin{align}
\textsc{tbabs} (
\textsc{ztbabs} \times \textsc{cflux} \times \textsc{zpowerlaw} +\textsc{apec})
\label{eq:pl_apec}
\end{align}
This model describes the spectrum of source no. 6 much better than does the PL model. To determine the statistical significance of the improvement of the fitting quality when adding the soft component to the absorbed power-law continuum, we calculated the ratio of the corresponding likelihoods. According to the Wilks theorem, $-2(\ln L_1 - \ln L_2)$ asymptotically converges to the $\chi^2$ distribution with the number of degrees of freedom equal to the difference of the numbers of degrees of freedom of two models one of which is embedded in the other. In our case, $-2\ln L_1$ and $-2\ln L_2$ correspond to the values of $Cstat$ for the model with and without the soft component, respectively. As a result, we find that for source no. 6 the probability that the improvement of the fitting quality when adding the soft component occurred by chance is $\sim 10^{-11}$. At the same time, it is important to note that our choice of the APEC model to describe the soft component in the spectrum, of course, is not unequivocal.

\subsubsection{The soft X-ray excess in the spectrum of SRGA\,J000132.9$+$240237/SRGeJ000132.4+24022}

Source no. 1 (Sy1.9), whose spectrum looks very hard compared to the remaining ones, stands out among the entire sample. When it is described by the PL model, the spectral slope turns out to be $\Gamma<0.5$ (see Table~\ref{tab:xray_params}), which differs radically from $\Gamma\sim 1.8$ typical for AGNs. Moreover, such hard power-law spectra cannot be obtained as a result of Comptonization in a hot plasma (see, e.g., \citealt{pozdnyakov83}).

Our attempts to ``correct'' the slope of the powerlaw component by adding the soft component have failed. In the case of SRGA\,J000132.9+240237, we may have to do with a strongly absorbed AGN, with the radiation reflected from the dusty torus making a major contribution to its spectrum. To describe the reflected component, we attempted to use the PEXRAV model \citep{pexrav}. Although this model is initially designed to describe the reflection of a power-law continuum from a flat surface of an optically thick neutral medium, it must describe approximately correctly the spectral shape of the reflected component for other geometries as well, in particular, in the case where a power-law continuum is reflected from a dusty torus (see, e.g., \citealt{melazzini23}).

In the language of XSPEC the resulting two-component model looks as follows (hereafter the PL+PEXRAV model):
\begin{align}
\textsc{tbabs} (\textsc{ztbabs} \times \textsc{zpowerlaw}\nonumber + \textsc{pexrav}).
\label{eq:pl_refl}
\end{align}
The normalization coefficient $rel_{\rm refl}$ of the PEXRAV model was specified to be negative to leave only the contribution of the reflected radiation in this compo- nent (since the direct component has already been taken into account in \textsc{zpowerlaw}). The following assumptions were made with regard to other parameters of the PEXRAV model: (1) the spectral slope of the incident continuum was tied to the slope of the \textsc{zpowerlaw} component, (2) no high-energy cutoff of the spectrum was introduced, (3) the chemical composition of the reflecting medium was fixed at its solar values, (4) the cosine of the angle of incidence of the radiation was taken to be 0.5, and (5) the normalization was tied to the normalization of the \textsc{zpowerlaw} component. All these assumptions have virtually no effect on the quality of the spectral fit.

Unfortunately, because of the insufficient statistics (25.7 photons minus the background in the \ero\ data), we failed to obtain any useful constraints on the parameters of the PL+PEXRAV model. On the other hand, as has already been said above, the slope $\Gamma< 0.5$ obtained in the PL model suggests that the radiation reflected from the dusty torus dominates in the spectrum. Therefore, we simplified the PL+PEXRAV model: the spectral slope was fixed at the ``canonical'' value of $\Gamma=1.8$ for AGNs, while the component describing the direct radiation was excluded. Within the standard AGN model this corresponds to the orientation at which the direct radiation is completely hidden by the torus. Such a model (PEXRAV) is described by the following expression:

\begin{equation}
\textsc{TBabs}\times\textsc{cflux}\times\textsc{pexrav}.
\label{eq:refl}
\end{equation}
According to the derived values of $Cstat$, this model describes the spectrum of source no. 1 as well as does the PL model.

To better study the properties of this interesting AGN, it is necessary to take an X-ray spectrum with good statistics in a wide energy range, desirably up to $\sim 50$--100~keV. This will allow one to use more physically justified models to describe the radiative transfer within the dusty torus model.

\subsection{Comments on Individual Objects}
\label{s:objects}

\subsubsection{№1. SRGA\,J000132.9$+$240237}

A powerful radio source that has been previously considered as a candidate for blazars \citep{dabrusco19} is associated with this X-ray source. In the optical spectrum investigated by us, which was taken during SDSS, there are narrow emission lines, while the H$\alpha$ line also has a broad component. This allows the object to be classified as a Seyfert 1.9 galaxy. Taking into account the very hard X-ray spectrum of the source (see above), it can be concluded that SRGA\,J000132.9$+$240237 is a narrow-line radio galaxy with strong absorption.

\subsubsection{№2. SRGA\,J001059.5$+$424341}

This X-ray source discovered during the \srg/\art\ all-sky survey did not enter into the published \ass\ catalog \citep{sazonov2024}, since it turned out to be slightly below the threshold established for this catalog in detection significance. However, since at the early stage of our work on the \ass\ catalog this source entered into the program of optical support for the survey and spectroscopic data were obtained, we included it in this paper. Through the accumulation of new \srg/\art\ data the source SRGA\,J001059.5$+$424341 may enter into the next official version of the catalog that is expected to be published after the completion of the eighth \srg/\art\ all-sky survey in 2025.

\subsubsection{№3. SRGA\,J023800.1$+$193818}

This X-ray source was first detected during the \rosat\ all-sky survey (2RXS\,J023759.4+193802, \citealt{2rxs}). Its optical counterpart is the galaxy 2MASS\,J02375999+1938118 with the IR color $W1-W2=0.11\pm 0.03$. The IR color atypical of AGNs is apparently related to the relatively low luminosity of this source ($\sim 10^{42}$\, erg\,s$^{-1}$ in X-rays). As a result, the radiation from the stellar population of the galaxy dominates over the radiation from the active nucleus in the near and mid-IR ranges.
 
\subsubsection{№5. SRGA\,J040335.6$+$472440}

This X-ray source was first detected during the \rosat\ all-sky survey (2RXS\,J040337.5+472432).

\subsubsection{№6. SRGA\,J165143.2$+$532539}

This X-ray source was known even from the Einstein obser- vations (2E\,1650.6+5330, \citealt{2e}) and was also detected during the \rosat\ all-sky survey (2RXS\,J165144.7+532532, \citealt{2rxs}). Its optical counterpart is the galaxy SBS\,1650+535, for which the redshift has already been measured previ- ously: $z=0.0287$ \citep{carrasco97}; however, the object was not known as an AGN. We classified it as Sy1.9 based on its optical spectrum. 

\subsubsection{№8. SRGA\,J191628.1$+$711619}

This source was discovered during the \rosat\ all-sky survey (2RXS\,J191627.3+711610). It is also present in the catalog of sources detected during the \xmm\ slew survey (XMMSL2\,J191627.8+711616, \citealt{saxton2008,xmmsl2}). 

\subsubsection{№9. SRGA\,J194412.5$-$243619}

This source was first detected during the \xmm\ slew survey (XMMSL2\,J194412.4-243621).

\subsubsection{№10. SRGA\,J195226.6$+$380011}

This source was discovered during the \rosat\ all-sky survey
(2RXS\,J195225.6+380017). It is also present in
the catalog of sources detecting during the \xmm\ slew survey (XMMSL2 J195225.3+380027).

\subsubsection{№11. SRGA\,J201633.2$+$705525}

This is a new X-ray source detected during the \srg/\art\ all-sky survey. There are narrow emission lines at $z=0.25791\pm 0.00005$ as well as broad H$\beta$ and H$\gamma$ lines in its optical spectrum, which allows the object to be classified as Sy1.

\section{DISCUSSION}

\begin{table*}
\caption{
AGN properties
}
\label{tab:class}
\centering
\begin{tabular}{rlccc}
\toprule
№ & Object & Optical type & $z^1$ & $\log\lx^2$ \\
\midrule
1  & SRGA\,J000132.9$+$240237 &  Sy1.9 & $0.10478\pm0.00006$ & $43.6_{-0.6}^{+0.2}$\\
2  & SRGA\,J001059.5$+$424341 &  Sy1   & $0.14640\pm0.00009$ & $43.70_{-0.12}^{+0.08}$\\
3  & SRGA\,J023800.1$+$193818 &  Sy1   & $0.03350\pm0.00014$ & $42.32_{-0.08}^{+0.08}$\\
4  & SRGA\,J025900.3$+$502958 &  Sy1   & $0.09461\pm0.00013$ & $43.51_{-0.13}^{+0.13}$\\
5  & SRGA\,J040335.6$+$472440 &  Sy1   & $0.0967\pm0.0003$   & $43.73_{-0.15}^{+0.11}$\\
6  & SRGA\,J165143.2$+$532539 &  Sy1.9 & $0.02864\pm0.00004$ & $42.26_{-0.11}^{+0.12}$\\
7  & SRGA\,J181749.5$+$234311 &  Sy1.9 & $0.08134\pm0.00014$ & $43.85_{-0.12}^{+0.12}$\\
8  & SRGA\,J191628.1$+$711619 &  Sy1   & $0.09839\pm0.00017$ & $43.48_{-0.05}^{+0.05}$\\
9  & SRGA\,J194412.5$-$243619 &  Sy2   & $0.14021\pm0.00010$ & $44.40_{-0.15}^{+0.12}$\\
10 & SRGA\,J195226.6$+$380011 &  Sy1   & $0.07666\pm0.00007$ & $43.33_{-0.09}^{+0.09}$\\
11 & SRGA\,J201633.2$+$705525 &  Sy1   & $0.25791\pm0.00005$ & $44.07_{-0.12}^{+0.11}$\\
\bottomrule
\end{tabular}
\begin{flushleft}
    $^1$ The Redshift measured from narrow emission lines.
    
    $^2$ The luminosity of the power-law spectral component in the observed 2--10~keV energy band in units of erg s$^{-1}$ corrected for the Galactic and intrinsic absorption based on the PL+APEC model for source no. 6 and the PL model for all the remaining ones (see Table~\ref{tab:xray_params}). The errors in the redshift are given at 68\% confidence, while those in the luminosity are given at 90\% confidence (without including the small redshift measurement error).
\end{flushleft}
\end{table*}

Table~\ref{tab:class} presents some physical characteristics of the AGNs investigated in this paper: the optical class, the redshift, and the X-ray luminosity $\lx$. The latter was calculated for the power-law component in the X-ray spectrum (for the model parameters from Table~\ref{tab:xray_params}) in the observed 2--10~keV energy band (taking into account the small redshifts of the objects, the $k$-correction was ignored) and was corrected for the Galactic and intrinsic absorption.

\begin{figure}
\centering
\includegraphics[width=1\columnwidth]{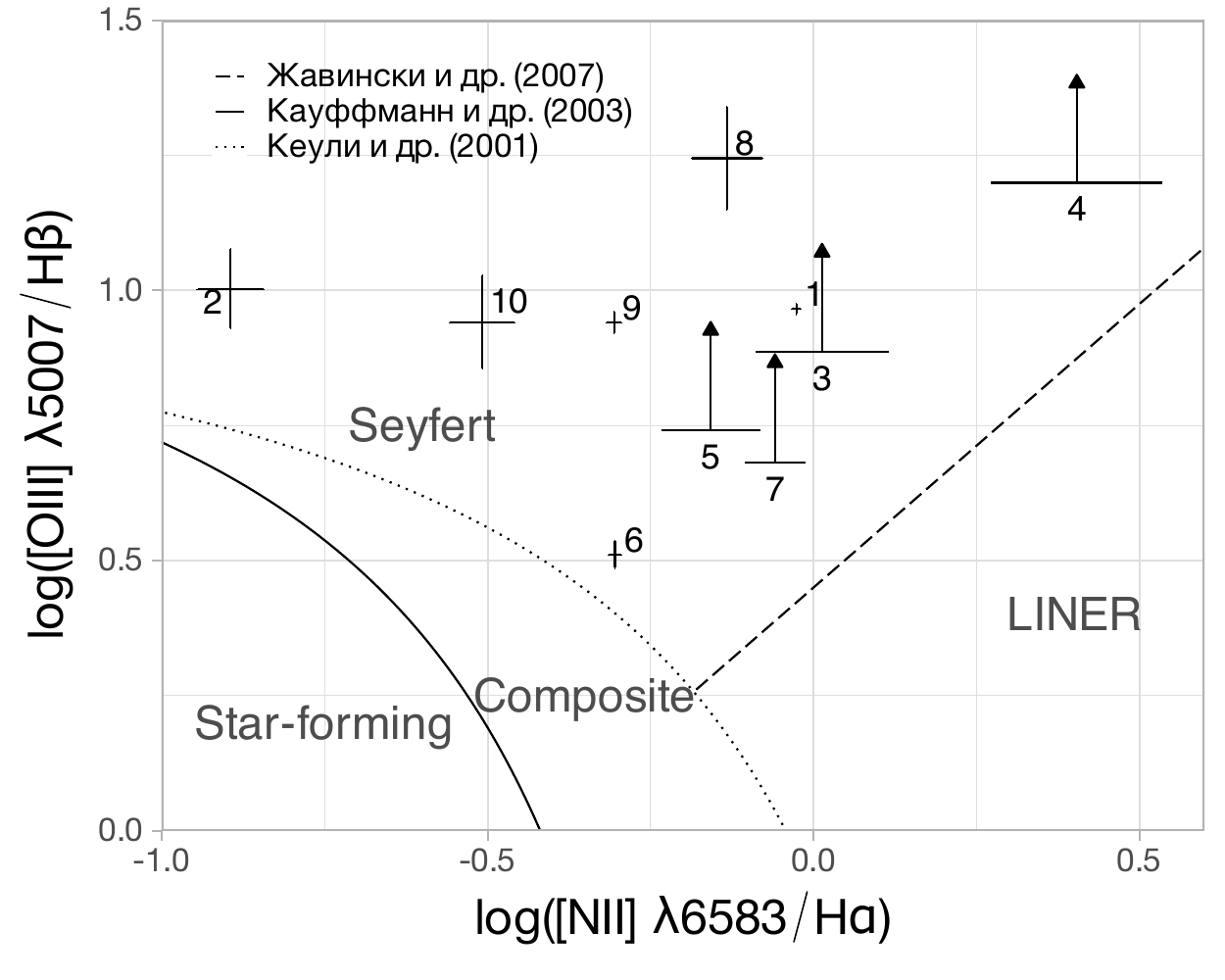}
\caption{Positions of the AGNs being studied on the BPT diagram \protect\citep{bpt}. The 1$\sigma$ confidence intervals for the flux ratios are shown. The arrows indicate the 2$\sigma$ lower limits. The demarcation lines between different classes of galaxies were taken from the following papers: \protect\cite{kauff03} -- the solid line, \protect\cite{kewly01} -- the dotted line, and \protect\cite{scha07} -- the dashed line. The sources are marked by the numbers given in Table~\ref{tab:sources}. Source no. 11 (SRGA\,J201633.2$+$705525) did not fall on the diagram, since the \protect\ha\ region for it was outside the spectral range.
}
\label{chart:bpt}
\end{figure}

The X-ray luminosities of the objects vary in the range from $\sim 2\times 10^{42}$ to $\sim 3\times 10^{44}$\,erg s$^{-1}$. Such values are typical for AGNs at the present epoch (see, e.g., \citealt{ueda14,sazonov15}). According to the narrow-line flux ratios \n2ha and \o3hb, all of the sources except object no. 11 (for which part of the necessary information is absent) fall into the region of Seyfert galaxies on the Baldwin–Phillips–Terlevich (BPT) diagram \citep{bpt} shown in Fig.~\ref{chart:bpt}.

\begin{figure}
  \centering
    \includegraphics[width=1\columnwidth]{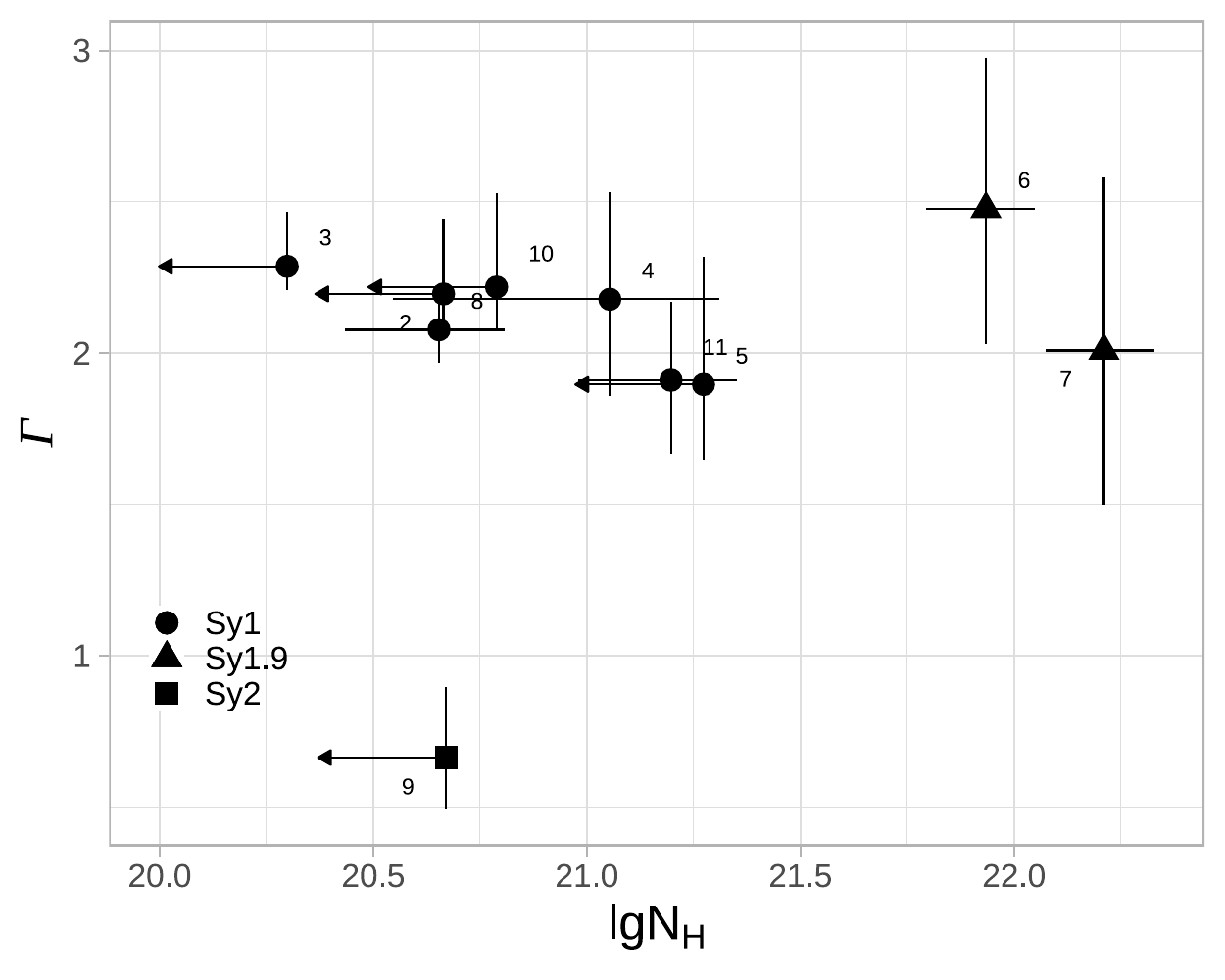}
    \caption{
    Slope of the X-ray power-law continuum versus intrinsic absorption column density. Seyfert galaxies of different types are indicated by different symbols. The errors and upper limits correspond to the 90\% confidence intervals. We used the PL+APEC model for source no. 6 and the PL model for the remaining ones (see Table~\ref{tab:xray_params}). Source no. 1 is not shown, since we failed to reliably constrain the parameters $\Gamma$ and $\nh$ for it.
    } 
  \label{chart:xray}
\end{figure}

In Fig.~\ref{chart:xray} the slope of the power-law continuum $\Gamma$ is plotted against the intrinsic absorption $\nh$ for the objects being studied. Most of the slopes are close, within the error limits, to $\Gamma\approx 1.8$ typical for AGNs (see, e.g., \citealt{trakhtenbrot17}). A statistically significant, but moderate ($\nh\sim 10^{22}$\,cm$^{-2}$) intrinsic absorption was revealed only in two objects (nos. 6 and 7, both are Sy1.9). One more source (no. 1, also Sy1.9) may be a strongly absorbed AGN, as discussed above; however, we fail to reliably constrain the parameters $\Gamma$ and $\nh$ for it based on the available data from the \srg\ all-sky survey. We found no evidence of significant intrinsic absorption ($\nh\lesssim 2\times 10^{21}$\,cm$^{-2}$) in the X-ray spectra of all seven Seyfert 1 galaxies and the Seyfert 2 galaxy (no. 9).

For Seyfert 1 galaxies we can estimate the masses of their central black holes from the luminosity and the width of the broad \ha\ or \hb\ emission line (the latter is relevant for object no. 11 in our sample) based on the well-known empirical relationship (Eqs. (6) and (7) in \citealt{green2005}) using the line fluxes and widths from Table~\ref{tab:lines}. The masses derived in this way lie in the range from $\sim 5\times 10^6$ to $\sim 1.5\times 10^8\,M_{\odot}$. In addition, we can estimate the bolometric luminosities and the accretion rates with respect to the critical one ($\lambda_{\rm Edd}$) using a typical bolometric correction for the 2--10\,keV energy band: $L_{\rm bol}/\lx\sim 11$ \citep{sazonov2012}. Our estimates are given in Table~\ref{tab:masses}. It is important to note that in addition to the specified statistical errors, there is a more significant systematic uncertainty associated with the determination of the black hole masses and the application of the bolometric correction. The values of $\lambda_{\rm Edd}$ obtained vary from  $\sim 3$\% to $\sim 20$\%, typical for Seyfert galaxies as a whole (see, e.g., \citealt{prokhorenko21,ananna2022}).

\begin{table*}
\caption{
The masses, bolometric luminosities, and Eddington ratios for the central black holes in Sy1 galaxies.
}
\label{tab:masses}
\centering
\begin{tabular}{rcccc}
\toprule
№ & Object & BH mass, $10^6 M_\odot$ & $L_{\rm bol}$, $10^{44}$ erg s$^{-1}$ & $\lambda_{\rm Edd}$\\
\midrule
2 & SRGAJ001059.5+424341 & $150\pm32$ & $5.5_{-1.3}^{+1.2}$ & $0.028\pm0.009$\\
3 & SRGAJ023800.1+193818 & $4.68\pm0.98$ & $0.23_{-0.04}^{+0.05}$ & $0.038\pm0.011$\\
4 & SRGAJ025900.3+502958 & $72\pm14$ & $3.6_{-0.9}^{+1.2}$ & $0.038\pm0.014$\\
5 & SRGAJ040335.6+472440 & $28\pm6$ & $5.8_{-1.7}^{+1.7}$ & $0.16\pm0.06$\\
8 & SRGAJ191628.1+711619 & $102\pm21$ & $3.3_{-0.4}^{+0.4}$ & $0.025\pm0.006$\\
10 & SRGAJ195226.6+380011 & $16\pm3$ & $2.3_{-0.4}^{+0.5}$ & $0.12\pm0.03$\\
11 & SRGAJ201633.2+705525 & $137\pm20$ & $13_{-3}^{+4}$ & $0.07\pm0.02$\\
\bottomrule
\end{tabular}
\end{table*}

\section{CONCLUSIONS}

Using the observations carried out with the \azt\ telescope and the archival spectroscopic data from SDSS and 6dF, we identified 11 new AGNs among the X-ray sources detected in the 4--12~keV energy band during the first five \srg\/\art\ all-sky surveys. All these sources are also detected with confidence by \ero\ in the 0.2--8.0~keV energy band; based on its data, we obtained a more accurate localization of the sources and took and analyzed their X-ray spectra. All these the objects turned out to be nearby Seyfert galaxies (7 Sy1, 3 Sy1.9, 1 Sy2) at redshifts $z = 0.029$--0.258.

Our analysis of the \ero\ X-ray spectra revealed a noticeable intrinsic absorption ($\nh\sim 10^{22}$\,cm$^{-2}$) in two of the four Seyfert 2 galaxies (Sy1.9, Sy2). The spectrum of one more of them (SRGA\,J000132.9+240237) is described by a power law with a slope $\Gamma<0.5$, which may point to a strong absorption and a significant contribution of the radiation reflected from the dusty torus. However, the available \srg\ all-sky survey data are not enough to obtain reliable constraints on the absorption column density in this object, which is also interesting in that it is radio loud. We are planning to take a higher-quality X-ray spectrum of this source in the pointing mode with the \srg/\art\ telescope to study in detail its physical properties.

This paper continues our series of publications on the optical identification of X-ray sources detected during the \srg/\art\ all-sky survey. Obtaining a large, statistically complete sample of AGNs selected by their emission in the hard 4--12~keV X-ray energy band must become the result of this work.

\section*{Acknowledgements}
In this study we used data from the \art\ and \ero\ telescopes onboard the \srg\ observatory. The \srg\ observatory was designed by the Joint-Stock Company ``Lavochkin Research and Production Association'' (within the Roskosmos State Corporation) with the participation of the Deutsches Zentrum fur Luft- und Raumfahrt (DLR) within the framework of the Russian Federal Space Program at the order of the Russian Academy of Sciences. The \ero\ X-ray telescope was built by a consortium of German Institutes led by MPE, and supported by DLR. The \art\ team thanks the Roskosmos State Corporation, the Russian Academy of Sciences, and the Rosatom State Corporation for supporting the \art\ design and production and the Joint-Stock Company ``Lavochkin Research and Production Association'' and its partners for the production and work with the spacecraft and the ``Navigator'' platform. The \ero\ data used in this paper were processed with the \textit{eSASS} software developed by the German eROSITA consortium and the proprietary data reduction and analysis software developed by the Russian eROSITA Consortium. This study was supported by RSF grant no. 19-12-00396.

\bibliographystyle{mnras}
\bibliography{pazh_bib_utf8}

\section{APPENDIX}

\begin{table*}
\caption{
\label{tab:lines} Spectral features of the sources. The wavelengths are in the observer’s frame. The fluxes, equivalent widths, and $FWHM$ were obtained for the reference frame of the sources. The confidence intervals are given at 1$\sigma$, while the upper limits are given at 2$\sigma$}
\centering
\begin{tabular}[t]{lcccc}
\toprule
Line & Wavelength & Flux, $10^{-15}$erg s$^{-1}$ cm$^{-2}$ & EW, \AA & FWHM, $10^2$ km s$^{-1}$\\
\midrule
\addlinespace[0.3em]
\multicolumn{5}{c}{SRGA\,J000132.9+240237}\\
{}[NeV]3346 & 3696 & $0.60\pm0.05$ & $-5.6\pm0.5$ & $-$\\
{}[NeVI]3426 & 3784 & $1.57\pm0.05$ & $-15.6\pm0.5$ & $-$\\
{}[OII]3726 & 4118 & $4.68\pm0.06$ & $-45.7\pm0.6$ & $-$\\
HeI 3888 & 4296 & $0.31\pm0.02$ & $-2.7\pm0.2$ & $-$\\
{}[SII]4071 & 4495 & $0.19\pm0.02$ & $-1.47\pm0.18$ & $-$\\
\hd & 4531 & $0.38\pm0.02$ & $-2.82\pm0.18$ & $-$\\
\hg & 4795 & $1.31\pm0.04$ & $-8.3\pm0.2$ & $-$\\
{}[OIII]4363 & 4818 & $1.44\pm0.04$ & $-8.9\pm0.3$ & $-$\\
\hb & 5370 & $1.69\pm0.04$ & $-7.26\pm0.15$ & $-$\\
{}[OIII]4959 & 5478 & $5.53\pm0.06$ & $-23.5\pm0.2$ & $-$\\
{}[OIII]5007 & 5531 & $15.62\pm0.12$ & $-66.1\pm0.5$ & $-$\\
{}[OI]6300 & 6961 & $1.07\pm0.04$ & $-4.00\pm0.14$ & $-$\\
{}[NII]6548 & 7235 & $3.74\pm0.07$ & $-13.7\pm0.3$ & $-$\\
\ha & 7250 & $8.29\pm0.09$ & $-30.2\pm0.3$ & $-$\\
{}[NII]6583 & 7273 & $7.80\pm0.09$ & $-28.4\pm0.3$ & $-$\\
? & 7310 & $23.6\pm0.3$ & $-85.5\pm0.9$ & $68.8\pm0.7$\\
{}[SII]6716 & 7420 & $1.78\pm0.05$ & $-6.41\pm0.16$ & $-$\\
{}[SII]6731 & 7437 & $1.56\pm0.05$ & $-5.61\pm0.16$ & $-$\\
\addlinespace[0.3em]
\multicolumn{5}{c}{SRGA\,J001059.5+424341}\\
{}[OII]3726 & 4274 & $1.9\pm0.2$ & $-9.3\pm1.1$ & $-$\\
\hg & 4976 & $0.65\pm0.11$ & $-3.8\pm0.6$ & $-$\\
{}[OIII]4363 & 5001 & $0.80\pm0.11$ & $-4.7\pm0.6$ & $-$\\
\hb, broad & 5572 & $10.7\pm0.5$ & $-58\pm3$ & $68\pm3$\\
\hb & 5572 & $0.94\pm0.16$ & $-5.1\pm1.0$ & $-$\\
{}[OIII]4959 & 5685 & $2.92\pm0.11$ & $-16.0\pm0.6$ & $-$\\
{}[OIII]5007 & 5740 & $9.44\pm0.15$ & $-53.0\pm0.8$ & $-$\\
{}[OI]6300 & 7226 & $0.36\pm0.07$ & $-2.7\pm0.5$ & $-$\\
{}[NII]6548 & 7509 & $0.11\pm0.03$ & $-0.8\pm0.2$ & $-$\\
\ha, broad & 7523 & $49.0\pm0.6$ & $-369\pm4$ & $61.7\pm0.9$\\
\ha & 7523 & $2.66\pm0.15$ & $-20.0\pm1.1$ & $-$\\
{}[NII]6583 & 7544 & $0.34\pm0.03$ & $-2.5\pm0.7$ & $-$\\
{}[SII]6716 & 7700 & $0.47\pm0.07$ & $-3.6\pm0.5$ & $-$\\
{}[SII]6731 & 7716 & $0.55\pm0.11$ & $-4.2\pm0.8$ & $-$\\
\addlinespace[0.3em]
\multicolumn{5}{c}{SRGA\,J023800.1+193818}\\
\hb, broad & 5029 & $6.2\pm0.8$ & $-17\pm2$ & $44\pm5$\\
\hb & 5029 & $ < 0.3$ & $ > -0.7$ & $-$\\
{}[OIII]4959 & 5125 & $0.90\pm0.14$ & $-2.4\pm0.4$ & $-$\\
{}[OIII]5007 & 5175 & $2.56\pm0.17$ & $-6.8\pm0.5$ & $-$\\
{}[NII]6548 & 6768 & $0.29\pm0.06$ & $-0.64\pm0.10$ & $-$\\
\ha, broad & 6783 & $21.2\pm0.6$ & $-48.8\pm1.4$ & $32.8\pm1.0$\\
\ha & 6783 & $0.84\pm0.19$ & $-1.9\pm0.4$ & $-$\\
{}[NII]6583 & 6803 & $0.86\pm0.06$ & $-1.9\pm0.3$ & $-$\\
{}[SII]6716 & 6942 & $ < 0.2$ & $ > -1.2$ & $-$\\
{}[SII]6731 & 6956 & $0.38\pm0.11$ & $-0.9\pm0.2$ & $-$\\
\addlinespace[0.3em]
\bottomrule
\end{tabular}
\end{table*}

\addtocounter{table}{-1}

\begin{table*}
\caption{(Contd.)}
\centering
\begin{tabular}{lcccc}
\toprule
Line & Wavelength & Flux, $10^{-15}$erg s$^{-1}$ cm$^{-2}$ & $EW$, \AA & FWHM, $10^2$ km s$^{-1}$\\
\midrule
\addlinespace[0.3em]
\multicolumn{5}{c}{SRGA\,J025900.3+502958}\\
\hg, broad & 4753 & $19\pm3$ & $-26\pm3$ & $48\pm3$\\
\hb, broad & 5322 & $35\pm2$ & $-42\pm3$ & $43\pm3$\\
\hb & 5322 & $ < 1.5$ & $ > -3$ & $-$\\
{}[OIII]4959 & 5429 & $8.7\pm0.6$ & $-10.5\pm0.7$ & $-$\\
{}[OIII]5007 & 5481 & $23.8\pm0.7$ & $-28.8\pm0.8$ & $-$\\
{}[NII]6548 & 7165 & $2.1\pm0.2$ & $-2.7\pm0.3$ & $-$\\
\ha, broad & 7180 & $153\pm2$ & $-197\pm3$ & $40.9\pm0.6$\\
\ha & 7180 & $2.4\pm0.7$ & $-3.1\pm1.1$ & $-$\\
{}[NII]6583 & 7210 & $6.2\pm0.2$ & $-8.1\pm0.8$ & $-$\\
{}[SII]6716 & 7352 & $2.7\pm0.3$ & $-3.6\pm0.3$ & $-$\\
{}[SII]6731 & 7370 & $2.7\pm0.3$ & $-3.6\pm0.3$ & $-$\\
\addlinespace[0.3em]
\multicolumn{5}{c}{SRGA\,J040335.6+472440}\\
\hb, broad & 5340 & $27\pm2$ & $-59\pm5$ & $33\pm3$\\
\hb & 5340 & $ < 1.9$ & $ > -4$ & $-$\\
{}[OIII]4959 & 5437 & $3.9\pm0.6$ & $-8.7\pm1.3$ & $-$\\
{}[OIII]5007 & 5489 & $10.4\pm0.8$ & $-23.3\pm1.7$ & $-$\\
{}[NII]6548 & 7183 & $2.2\pm0.4$ & $-6.7\pm1.4$ & $-$\\
\ha, broad & 7198 & $60\pm3$ & $-185\pm10$ & $32.7\pm1.4$\\
\ha & 7198 & $9.5\pm1.5$ & $-30\pm6$ & $-$\\
{}[NII]6583 & 7219 & $6.6\pm0.4$ & $-20\pm4$ & $-$\\
\addlinespace[0.3em]
\multicolumn{5}{c}{SRGA\,J165143.2+532539}\\
\hb & 5001 & $29.8\pm1.6$ & $-6.6\pm0.3$ & $-$\\
{}[OIII]4959 & 5100 & $34.2\pm1.5$ & $-7.7\pm0.3$ & $-$\\
{}[OIII]5007 & 5149 & $96.7\pm1.9$ & $-21.8\pm0.4$ & $-$\\
{}[OI]6300 & 6482 & $8.0\pm1.0$ & $-1.9\pm0.2$ & $-$\\
{}[NII]6548 & 6736 & $23.4\pm0.7$ & $-5.82\pm0.19$ & $-$\\
\ha, broad & 6751 & $97\pm5$ & $-24.6\pm1.3$ & $30.6\pm1.7$\\
\ha & 6751 & $141\pm3$ & $-35.4\pm1.0$ & $-$\\
{}[NII]6583 & 6772 & $70.1\pm0.7$ & $-17.6\pm0.6$ & $-$\\
{}[SII]6716 & 6909 & $21.0\pm1.0$ & $-5.5\pm0.2$ & $-$\\
{}[SII]6731 & 6924 & $20.5\pm1.1$ & $-5.4\pm0.3$ & $-$\\
\addlinespace[0.3em]
\multicolumn{5}{c}{SRGA\,J181749.5+234311}\\
\hb & 5257 & $ < 2$ & $ > -4$ & $-$\\
{}[OIII]4959 & 5358 & $3.7\pm1.1$ & $-4.9\pm1.3$ & $-$\\
{}[OIII]5007 & 5414 & $11.2\pm1.3$ & $-13.8\pm1.5$ & $-$\\
{}[OI]6300 & 6813 & $7.2\pm0.9$ & $-8.1\pm1.0$ & $-$\\
{}[NII]6548 & 7082 & $4.0\pm0.4$ & $-4.4\pm0.5$ & $-$\\
\ha, broad & 7097 & $43\pm5$ & $-47\pm5$ & $54\pm7$\\
\ha & 7097 & $13.6\pm1.4$ & $-14.8\pm1.5$ & $-$\\
{}[NII]6583 & 7118 & $11.9\pm0.4$ & $-13.0\pm1.5$ & $-$\\
\addlinespace[0.3em]
\bottomrule
\end{tabular}
\end{table*}

\addtocounter{table}{-1}

\begin{table*}
\caption{(Contd.)}
\centering
\begin{tabular}{lcccc}
\toprule
Line & Wavelength & Flux, $10^{-15}$erg s$^{-1}$ cm$^{-2}$ & $EW$, \AA & FWHM, $10^2$ km s$^{-1}$\\
\midrule
\addlinespace[0.3em]
\multicolumn{5}{c}{SRGA\,J191628.1+711619}\\
{}[OII]3726 & 4096 & $24\pm4$ & $-10.8\pm1.8$ & $-$\\
\hg, broad & 4778 & $30\pm4$ & $-18.1\pm1.8$ & $47\pm3$\\
\hb, broad & 5339 & $58\pm4$ & $-31.0\pm1.9$ & $42\pm2$\\
\hb & 5339 & $5.1\pm1.1$ & $-2.8\pm0.5$ & $-$\\
{}[OIII]4959 & 5447 & $28.7\pm0.9$ & $-16.0\pm0.5$ & $-$\\
{}[OIII]5007 & 5500 & $90.1\pm1.3$ & $-52.0\pm0.7$ & $-$\\
{}[OI]6300 & 6923 & $2.7\pm0.6$ & $-1.9\pm0.4$ & $-$\\
{}[NII]6548 & 7193 & $2.9\pm0.4$ & $-2.1\pm0.2$ & $-$\\
\ha, broad & 7208 & $230\pm4$ & $-170\pm3$ & $42.5\pm0.8$\\
\ha & 7208 & $11.8\pm1.4$ & $-8.6\pm0.8$ & $-$\\
{}[NII]6583 & 7228 & $8.7\pm0.4$ & $-6.4\pm0.7$ & $-$\\
\addlinespace[0.3em]
\multicolumn{5}{c}{SRGA\,J194412.5-243619}\\
{}[OII]3726 & 4251 & $-$ & $-24.9\pm0.7$ & $-$\\
\hg & 4950 & $-$ & $-5.1\pm0.5$ & $-$\\
{}[OIII]4363 & 4974 & $-$ & $-5.1\pm0.4$ & $-$\\
\hb & 5544 & $-$ & $-9.9\pm0.5$ & $-$\\
{}[OIII]4959 & 5654 & $-$ & $-29.9\pm0.7$ & $-$\\
{}[OIII]5007 & 5709 & $-$ & $-85.7\pm0.9$ & $-$\\
{}[OI]6300 & 7184 & $-$ & $-5.3\pm0.7$ & $-$\\
{}[NII]6548 & 7468 & $-$ & $-5.05\pm0.16$ & $-$\\
\ha & 7483 & $-$ & $-30.7\pm0.7$ & $-$\\
{}[NII]6583 & 7503 & $-$ & $-15.1\pm0.5$ & $-$\\
\addlinespace[0.3em]
\multicolumn{5}{c}{SRGA\,J195226.6+380011}\\
\hg, broad & 4683 & $6.2\pm0.6$ & $-23\pm2$ & $34\pm2$\\
\hb, broad & 5234 & $8.6\pm0.6$ & $-22.2\pm1.5$ & $31\pm2$\\
\hb & 5234 & $1.3\pm0.3$ & $-3.4\pm0.7$ & $-$\\
{}[OIII]4959 & 5339 & $4.0\pm0.2$ & $-10.0\pm0.5$ & $-$\\
{}[OIII]5007 & 5390 & $11.5\pm0.3$ & $-28.4\pm0.6$ & $-$\\
{}[NII]6548 & 7052 & $0.81\pm0.15$ & $-1.8\pm0.4$ & $-$\\
\ha, broad & 7066 & $59.1\pm1.3$ & $-134\pm4$ & $28.2\pm0.5$\\
\ha & 7066 & $7.8\pm0.8$ & $-18\pm3$ & $-$\\
{}[NII]6583 & 7087 & $2.43\pm0.15$ & $-5.5\pm1.2$ & $-$\\
\addlinespace[0.3em]
\multicolumn{5}{c}{SRGA\,J201633.2+705525}\\
{}[OII]3726 & 4688 & $1.5\pm0.4$ & $-4.3\pm1.0$ & $-$\\
\hg, broad & 5479 & $7.0\pm0.7$ & $-28\pm3$ & $55\pm4$\\
\hb, broad & 6115 & $10.9\pm0.7$ & $-48\pm4$ & $49\pm3$\\
\hb & 6115 & $1.36\pm0.18$ & $-6.1\pm0.9$ & $-$\\
{}[OIII]4959 & 6238 & $1.92\pm0.18$ & $-8.4\pm0.8$ & $-$\\
{}[OIII]5007 & 6298 & $6.8\pm0.3$ & $-29.1\pm1.1$ & $-$\\
\addlinespace[0.3em]
\bottomrule
\end{tabular}
\end{table*}

\end{document}